\documentclass[
reprint,
superscriptaddress,
groupedaddress,
amsmath,amssymb,
aps,
prb,
]{revtex4-2}

\usepackage{amsmath}
\usepackage{amssymb}
\usepackage[bookmarks=true,colorlinks,citecolor=blue,linkcolor=blue,allcolors=blue]{hyperref}

\usepackage[capitalise]{cleveref}
\usepackage{physics}
\usepackage[toc,page]{appendix}
\usepackage{graphicx}
\usepackage{makecell}
\usepackage{calligra}

\usepackage{booktabs}       
\usepackage{makecell}

\usepackage{relsize}

\DeclareMathOperator{\spn}{\text{span}}

\DeclareMathOperator{\lbb}{[\![\!}
\DeclareMathOperator{\rbb}{\!]\!]}

\DeclareMathOperator{\lwa}{\langle \! \langle \! }
\DeclareMathOperator{\rwa}{ \! \rangle \! \rangle_\wedge }

\DeclareMathOperator{\Ex}{ \mathbb{E} }
\DeclareMathOperator{\Var}{ \text{Var} }
\DeclareMathOperator{\Cov}{\text{Cov} }

\DeclareMathOperator{\QGT}{S}

\begin{document}

\title{Grassmann Variational Monte Carlo with neural wave functions}

\author{Douglas Hendry}
\affiliation{Institute of Physics, \'{E}cole Polytechnique F\'{e}d\'{e}rale de Lausanne (EPFL), CH-1015 Lausanne, Switzerland}
\affiliation{Center for Quantum Science and Engineering, EPFL, Lausanne, Switzerland}

\author{Alessandro Sinibaldi}
\affiliation{Institute of Physics, \'{E}cole Polytechnique F\'{e}d\'{e}rale de Lausanne (EPFL), CH-1015 Lausanne, Switzerland}
\affiliation{Center for Quantum Science and Engineering, EPFL, Lausanne, Switzerland}

\author{Giuseppe Carleo}
\affiliation{Institute of Physics, \'{E}cole Polytechnique F\'{e}d\'{e}rale de Lausanne (EPFL), CH-1015 Lausanne, Switzerland}
\affiliation{Center for Quantum Science and Engineering, EPFL, Lausanne, Switzerland}

\begin{abstract}
Excited states play a central role in determining the physical properties of quantum matter, yet their accurate computation in many-body systems remains a formidable challenge for numerical methods. 
While neural quantum states have delivered outstanding results for ground-state problems, extending their applicability to excited states has faced limitations, including instability in dense spectra and reliance on symmetry constraints or penalty-based formulations. 
In this work, we rigorously formalize the framework introduced by Pfau et al.~\cite{pfau2024accurate} in terms of Grassmann geometry of the Hilbert space.
This allows us to generalize the Stochastic Reconfiguration method for the simultaneous optimization of multiple variational wave functions, and to introduce the multidimensional versions of operator variances and overlaps. 
We validate our approach on the Heisenberg quantum spin model on the square lattice, achieving highly accurate energies and physical observables for a large number of excited states. 
\end{abstract}

\maketitle

\section{Introduction}
Quantum matter does not reveal its full character in the ground state. 
Optical spectra, thermodynamic response, and non-equilibrium dynamics are all controlled by the excited spectrum.
The exact computation of low-lying excited states suffers from the curse of dimensionality of quantum many-body Hilbert spaces, limiting exact calculations to a few particles.
For ab-initio problems, equation-of-motion coupled-cluster (EOM-CC) reduces the prefactor but not the exponential wall~\cite{Stanton1993}.
Quantum Monte Carlo helps in certain models, yet the sign problem usually comes back where correlations are strongest~\cite{Troyer2005}.
Tensor networks (TN) go furthest in one dimension: the density-matrix renormalization group (DMRG) gives almost exact spectra for gapped chains~\cite{white1992density}. In two and three dimensions — or in frustrated lattices — the relevant excitations carry more entanglement; higher-dimensional TN, such as PEPS~\cite{Verstraete2004} and MERA~\cite{Vidal2007}, treat only small clusters before the cost again balloons with system width and bond dimension.
Accurate excited-state calculations for higher-dimensional or frustrated quantum Hamiltonians, therefore, remain an open challenge.

Since their introduction, neural quantum states (NQS)~\cite{carleo2017solving} have provided several state-of-the-art results for ground state search, including spin~\cite{choo2019two,viteritti2023transformer_1d,viteritti2023transformer_2d,chen2024empowering}, bosonic~\cite{Denis2025accurateneural} and electronic models~\cite{Choo2020Fermions,Pfau2020PRR,NomuraPRBFermionic,Stokes2020PRB,Nys22Fermions}, as well as competitive performance in simulating quantum many-body dynamics~\cite{schmitt2020quantum, sinibaldi2023unbiasing,nys2024ab,sinibaldi2024time}. 
Extending that success to excited states is challenging. 
Early NQS schemes enforced point-group or momentum quantum numbers~\cite{choo_excited} but could only target a single eigenvector at a time. 
Penalty-based and folded-spectrum approaches~\cite{entwistle2023electronic,pathak2021excited} introduce auxiliary cost terms in the variational problem to discourage collapse into the ground state, at the price of complicated hyperparameter tuning and increased variance. 
Recently, Pfau et al.~\cite{pfau2024accurate} reformulated the excitation search as a ground-state problem of an enlarged Hamiltonian, eliminating the need for imposing symmetries and orthogonalization.

In this work, we rigorously generalize the framework of Ref.~\cite{pfau2024accurate} by formulating it in geometric terms, making use of the Grassmann structure inherent in the Hilbert space.
This method avoids to invert the possibly ill-conditioned Gram matrix of the basis states, which plays the role of a multidimensional normalization factor. 
As such, it provides a natural extension of standard Variational Monte Carlo~\cite{mcmillan1965ground} to the concurrent optimization of multiple quantum states.
In particular, we generalize the Stochastic Reconfiguration~\cite{sorella1998green} method to optimize several variational wave functions simultaneously, and we introduce multidimensional formulations of operator variances and overlaps.
We demonstrate the effectiveness of our approach by applying it in second quantization on the 2D Heisenberg quantum spin model, achieving remarkable accuracy in both the energy and the spin structure factor for a substantial number of excited states.

The remainder of the paper is organized as follows.
Section 2 introduces the various representations of linear subspaces within the Hilbert space.
Section 3 extends the definitions of expectation value, variance, and overlap to Grassmannians.
Section 4 formalizes the variational geometry of linear subspaces and presents the Grassmann Variational Monte Carlo method.
Section 5 describes the neural network architecture employed and presents numerical results for the two-dimensional Heisenberg model.
Finally, Section 6 concludes the paper and discusses directions for future research.

\section{Representations of Linear Subspaces }

\subsection{Basis Representation}
\label{sec:basis_representation}
We consider a Hilbert space $\mathcal{H}$.
The space comprising all possible $N$-dimensional linear subspaces $\mathcal{V}$ of $\mathcal{H}$ forms a differential manifold which is the Grassmannian $\mathbf{Gr}_N(\mathcal{H})$.
One of the most interesting and studied aspects of Grassmann manifolds is the multitude of ways they can be parameterized and embedded in other spaces. 

The most straightforward way to parametrize a linear subspace is through a chosen basis. 
We define a linear subspace $\mathcal{V} \in \mathbf{Gr}_N(\mathcal{H})$ in terms of a given set of basis states $\{\phi_{1},\ldots,\phi_{N}\}$ as the ``basis representation". 
This representation will serve as our primary reference point from which all the other representations presented will be defined in relation.
To facilitate this, it is convenient to consider ordered bases. 
We denote a particular ordered basis, and more generally any collection of individual states, with a capital letter and the individual states with the indexed lowercase as
\begin{equation}
        \Phi = ( \phi_{1},\ldots,\phi_{N}) \in \mathcal{H}^{\otimes N}.
\end{equation}

For each basis $\Phi$, its Gram matrix, denoted as $G(\Phi)$, is the $N \times N$  overlap matrix between the basis pairs. 
Its dual basis, denoted as $\widetilde{\Phi}$, is the equivalent basis such that the relative overlaps are given as $\langle \widetilde{\phi}_i|\phi_j \rangle = \delta_{ij}$.
\subsection{Matrix Representation}
The matrix representation is an alternative representation of a linear subspace, which is more suitable for working with Grassmannians.  
We start by introducing the matrix analog of the bra-ket inner product. 
This corresponds to the overlap matrix of two tuples of states $ \Phi = (\phi_{1}, \ldots, \phi_{N})$ and $ \Psi = (\psi_{1}, \ldots, \psi_{N})$  which is expressed as the overlap between the ``bra matrix" $\lbb \Psi |$ and the ``ket matrix"  $| \Phi \rbb$:

\begin{equation}
  \lbb \Psi | \Phi \rbb = 
    \begin{bmatrix}
    \langle \psi_{1}|\phi_{1} \rangle & \dots & \langle\psi_{1}|\phi_{N}\rangle \\
    \vdots & \ddots & \vdots \\
    \langle \psi_{N}|\phi_{1} \rangle & \ldots & \langle \psi_{N}|\phi_{N} \rangle
    \end{bmatrix}.
\end{equation}

The row vectors of the $N \times |\mathcal{H}|$ bra matrix $\lbb \Psi|$ and the column vectors of the $ |\mathcal{H}| \times N$ ket matrix $| \Phi \rbb$ are given by the ordered bras of $\Psi$ and kets of $\Phi$ respectively.
The $N \times N$ overlap matrix $\lbb \Psi | \Phi \rbb$ is given by all the ordered overlaps between the basis states.

For a linear subspace $\mathcal{V}$, the equations for their bases can be expressed compactly in this matrix representation.
For a basis $\Phi$, the Gram matrix is then $G = \lbb \Phi | \Phi \rbb$ and the the bra matrix of 
its dual basis $\widetilde{\Phi}$ is given by the pseudo inverse of the basis ket matrix $\lbb \widetilde{\Phi}| = \left( | \Phi \rbb \right)^{(+)} = G^{-1} \lbb \Phi |$. 
Changes of basis can be expressed in terms of matrix multiplications for any non-singular matrix $X \in GL_N(\mathbb{C})$ as  $|\Phi \rbb \rightarrow |\Phi \rbb \cdot X$ and $\lbb \widetilde{\Phi}| \rightarrow X^{-1} \cdot \lbb \widetilde{\Phi}| $ for the corresponding dual basis. 
The equivalent basis representation of the Grassmann manifold $\mathbf{Gr}_N(\mathcal{H})$ can now be recast in the matrix representation as the orbits/equivalence classes of full rank ket matrices with respect to group action on the right by $GL_N(\mathbb{C})$~\cite{absil2004riemannian}.

\subsection{Wedge Product Representation}
Another useful representation for Grassmannians, the so-called wedge product representation, is generated by the Plücker embedding $ \mathbf{Gr}_N(\mathcal{H}) \rightarrow \mathbf{P}(\wedge^{N}\mathcal{H})$, where $\wedge$ denotes the wedge product. 
This allows each linear subspace $\mathcal{V}$ to be represented uniquely (up to an overall constant) as a single wave-function in the expanded anti-symmetric Hilbert space $\wedge^{N}\mathcal{H}$. This provides a blueprint for formulating the Grassmann version of Variational Monte Carlo (VMC) as a standard single state VMC in $\wedge^{N}\mathcal{H}$, as done by Pfau et al.~\cite{pfau2024accurate}.
Considering a basis $\Phi$ of $\mathcal{V}$, the single state representation of $\mathcal{V}$ is given by the wedge product of the individual basis states as

\begin{equation}
    |\Phi \rwa = \bigwedge_{i=1}^{N} | \phi_{i} \rangle = \sum_{ \sigma \in  \{ \text{perms} \} } \text{sgn}  (\sigma) \,  |\phi_{\sigma {(1)} )}\rangle|\phi_{\sigma {(2)} }\rangle \ldots, 
\end{equation}
where the sum runs over all the possible permutations of the indices. 
We will refer to the previous representation as the ``wedge product representation" of $\mathcal{V}$.
Changes of basis in the matrix representation $|\Phi \rbb \rightarrow |\Phi \rbb \cdot X$ translate into the wedge product representation as $|\Phi \rangle\!\rangle \rightarrow \det(X) \cdot|\Phi \rangle\!\rangle$. 
This ensures that all the wedge products of equivalent bases lie in the same ray of the projective space $\mathbf{P}(\wedge^{N}\mathcal{H})$.
The bras $\lwa \Phi |$, from the wedge product of the individual state bras, are normalized by a factor of $1/N!$. 
Then, the wedge product overlaps relate to the corresponding overlap matrices as the determinants, namely:
\begin{equation}
    \lwa \Psi|\Phi \rwa =   \text{det} \left(\,  [\![ \Psi|\Phi ]\!]  \, \right).
    \label{eq:wedge_inner_det}
\end{equation}

\subsection{Projector and Density Operator Representations}
Every linear subspace $\mathcal{V}$ has a unique operator representation as the orthogonal projector $\hat{P}_{\mathcal{V}}: \mathcal{H} \rightarrow \mathcal{V}$, of which projects the full Hilbert space $\mathcal{H}$ onto itself.
In terms of the orthogonal eigenstates, $\hat{P}_{\mathcal{V}}$ is defined so that the $N$ eigenstates of eigenvalue one collectively span $\mathcal{V}$.
The remaining eigenstates, all of eigenvalue zero, then collectively span $\mathcal{V}^\perp$ as the set of states that are orthogonal to all the states in $\mathcal{V}$. 
The orthogonal projector for $\mathcal{V}^\perp$ is then the complementary projector $\hat{P}_{\mathcal{V}^\perp}=1-\hat{P}_{\mathcal{V}}$, which equivalently projects out of $\mathcal{V}$.
The orthogonal projector $\hat{P}_{\mathcal{V}}$ can be constructed from any basis $\Phi$ of $\mathcal{V}$ as the sum of the outer products of each basis and dual basis pair.
In the matrix representation, this corresponds simply to the basis ket matrix times the dual basis bra matrix:
\begin{equation}
    \hat{P}_{\mathcal{V}} =\sum_{k=1}^N |\phi_k \rangle \langle \widetilde{\phi}_k |= |\Phi \rbb \lbb \widetilde{\Phi}|.
\end{equation}

The dimension of any linear subspace $\mathcal{V}$ can be expressed algebraically as the trace of its orthogonal projector $\hat{P}_{\mathcal{V}}$.
Each Grassmannian $\mathbf{Gr}_N(\mathcal{H})$ can then be expressed bijectively as the set of the orthogonal projectors whose traces equal $N$. Additionally, by rescaling an orthogonal projector $\hat{P}_{\mathcal{V}}$ by the constant factor $1/N$, it becomes a density operator.
This then gives an equivalent ``density operator representation" for each linear subspace $\mathcal{V} \in \mathbf{Gr}_N(\mathcal{H})$:
\begin{equation}
\hat{\rho}_{\mathcal{V}} = \frac{1}{N} \hat{P}_{\mathcal{V}}.
\end{equation}

The density operator  $\hat{\rho}_{\mathcal{V}}$ for each $\mathcal{V}$ can be defined directly so that its support is $\mathcal{V}$ and its weighting of states in $\mathcal{V}$ is uniform. This equivalent representation will provide another perspective for quantum applications.
In the following sections, when attempting to extend quantum definitions to linear subspaces $\mathcal{V}$, the proposed quantities can be formulated from the mixed state definitions for $\hat{\rho}_{\mathcal{V}}$.

\section{Quantum Grassmannians}

\subsection{Extending Quantum Definitions to Linear Subspaces}
The term quantum Grassmannian can be found elsewhere in the literature with very different definitions. However, we will use it to refer generally to the extension of quantum concepts to Grassmannians.
The central idea is to generalize standard quantum definitions — typically formulated for rays as the unique representations of physical states — to linear subspaces in the Hilbert space. 
As the projective Hilbert space $P(\mathcal{H})$ is the Grassmannian $\mathbf{Gr}_N(\mathcal{H})$ for $N=1$, this extension is quite natural and much of the work in Grassmann geometry is rooted in this perspective as an extension of projective geometry.

Though in theory all or most aspects of quantum mechanics could be extended to Grassmannians, in this work we will be concerned only with the extension of basic quantum values such as operator expectation, variance,  overlap/fidelity, and their practical numerical applications to standard quantum problems. The Grassmann extension for a quantum value is given essentially by promoting the scalar value to a basis-dependent $N \times N$ matrix, whose basis-independent eigenvalues form a set of $N$ individual quantum values or similarly defined values.
These Grassmann matrix analogs for scalar quantum values arise across a broad range of contexts.

\subsubsection{Promoting Scalars to Matrices}

The promotion of scalars to matrices is formulated functionally, such that a defined quantum value as scalar mapping of states $\phi \rightarrow f(\phi)$ is extended to analogous matrix mapping of bases $\Phi \rightarrow F(\Phi)$.
The scalar mapping of states, which inherently is a mapping of rays, is invariant under multiplication by a constant $ |\phi\rangle \rightarrow x | \phi \rangle $ to ensure the same value for all the states in the same ray.
This translates for bases, as representations of $\mathcal{V}$, to equivariance of the matrix mapping for changes of equivalent basis.
To ensure the eigenvalues are inherent to $\mathcal{V}$, the equivariance takes the form of similarity transformations, so that $|\Phi \rbb \rightarrow |\Phi \rbb \cdot X$ results in $F \rightarrow X^{-1} \cdot F \cdot X$.

The exact form of the matrix promotion $F(\Phi)$ of any given scalar quantum value $f(\phi)$ can be derived by following the same general set of procedures.
This is because any relevant quantum value is formed functionally as the composite of some algebraically simple function of a set of overlaps between pairs of states or with an operator in between.
Firstly, each of the composite overlaps is promoted to an $N \times N$ overlap matrix by replacing the single state bras and kets of $\phi$  with the bra and ket matrices respectively of $\Phi$.
By plugging the overlap matrices into the overlap function, algebraic scalar multiplication and division become matrix multiplication and multiplication by matrix inverse respectively.
Due to the non-commutativity of matrix multiplications, this leaves multiple definitions for each  possible ordering of the matrices.
Finally, the correct ordering is identified as the one that ensures the desired equivariance under similarity transformations.

\subsubsection{Principal Values and Bases}

The construction of the matrix analog for a given quantum value ensures that the matrix eigenvalues are a set of $N$ corresponding quantum values, which we will refer to as the ``principal values".
Each principal value $\lambda_i$ is the given quantum value or a similarly defined value for a corresponding state $u_i \in \mathcal{V}$. Collectively, these corresponding ``principal states" $U$ are always constrained to be linearly independent and thus form a unique (up to individual constants) ``principal basis."
The principal basis $U$ is determined as the basis that diagonalizes the matrices as $F(U) = \Lambda$. Practically, the principal values and the basis can be obtained from any basis $\Phi$ by an eigendecomposition $F(\Phi) = X \cdot \Lambda \cdot X^{-1 } $, where the eigenvector matrix $X$ provides the change of basis $|U \rbb = | \Phi \rbb \cdot X$.

The Grassmann extension of a given quantum value can now be expressed properly in terms of linear subspaces and independently of their basis representations.
Fundamentally, the single quantum value for rays is extended to a set of $N$ quantum values for linear subspaces, given by the principal values.
As an additional component, the given quantum value also selects a unique basis for each linear subspace, given by the principal basis.

\subsection{Operators}

\subsubsection{Operator Expectation Matrices}
We will refer to the matrix promotion of a pure state operator expectation value as an operator expectation matrix (OEM).
For a Hermitian operator $\hat{A}$ and a basis $\Phi$ of $\mathcal{V}$, the OEM $\widetilde{A}(\Phi)$ is given by the operator overlap matrix $A(\Phi) = \lbb \Phi| \hat{A} | \Phi \rbb$ normalized on the left by the inverse of the Gram matrix, namely $\widetilde{A}(\Phi) = G^{-1}(\Phi)\cdot A(\Phi)$. 
The normalization equivalently transforms the basis bras into dual basis bras. 
The OEM $\widetilde{A}(\Phi)$ is thus the operator overlap matrix for mixed dual basis bra and basis ket pairs:
\begin{equation}
    \widetilde{A}(\Phi) = [\![ \widetilde{\Phi} |\hat{A}| \Phi ]\!] = 
    \begin{bmatrix}
    \langle \widetilde{\phi}_{1}|\hat{A}|\phi_{1} \rangle & \dots & \langle\widetilde{\phi}_{1}|\hat{A}|\phi_{N}\rangle \\
    \vdots & \ddots & \vdots \\
    \langle \widetilde{\phi}_{N}|\hat{A}|\phi_{1} \rangle & \ldots & \langle \widetilde{\phi}_{N}|\hat{A}|\phi_{N} \rangle
    \end{bmatrix}.
\end{equation}

The OEMs of $\hat{A}$ for all the bases of $\mathcal{V}$ encode all the pure state expectation values for every state in $\mathcal{V}$. Transforming to any orthogonal basis $|\Phi\rbb \rightarrow |\Phi \rbb \cdot X$ with $\widetilde{A} \rightarrow X^{-1} \cdot \widetilde{A} \cdot X $, the diagonal matrix elements give each individual $\langle \hat{A} \rangle_{\phi_i} \equiv \bra{\phi_i} \hat{A} \ket{\phi_i} / \bra{\phi_i} \ket{\phi_i}$.
The principal operator expectation basis $U$ is then the basis for which all the off-diagonal entries are zero.
The principal operator expectation values $\{ \lambda_i \}$, as the OEM eigenvalues for every basis of $\mathcal{V}$, are each individually the pure state expectations $\lambda_i = \langle \hat{A} \rangle_{u_i}$.

The OEMs and the importance of the principal values and basis can be understood in terms of operator projections.
An operator $\hat{A}$ acting on the full Hilbert space $\mathcal{H}$ is projected to an operator acting only on $\mathcal{V}\subset \mathcal{H}$ as $\hat{A}_\mathcal{V} = \hat{P}_{\mathcal{V}} \hat{A} \hat{P}_{\mathcal{V}}$.
Expressed in terms of any basis $\Phi$, the matrix elements of the projected operator $\hat{A}_\mathcal{V}$ are then given by the OEM $\widetilde{A}(\Phi)$. As the principal basis OEM is diagonal $\widetilde{A}(U) = \Lambda$, they form the eigenstates of $\hat{A}_\mathcal{V}$ and the principal values are the eigenvalues.
Therefore, we can write:
\begin{equation}
\begin{aligned}
    \hat{A}_\mathcal{V} &= |\Phi \rbb \cdot \widetilde{A}(\Phi) \cdot \lbb  \widetilde{\Phi}| =  \sum_{ij} \widetilde{A}_{ij}(\Phi)\, |\phi_i\rangle\langle \widetilde{\phi}_j | \\
    &= |U \rbb \cdot \widetilde{A}(U) \cdot \lbb  \widetilde{U}| =  \sum_{k} \lambda_k \, |u_k\rangle\langle \widetilde{u}_k |.
\end{aligned} 
\end{equation}

As the eigenstates of $\hat{A}_\mathcal{V}$, the uniqueness (up to overall constants) of $U$ is guaranteed only if the principal values/eigenvalues are not degenerate.
Importantly, these degrees of freedom for $U$  correspond directly with the degrees for $X$ from eigendecomposition $\widetilde{A}(\Phi) = X \cdot \Lambda \cdot X^{-1}$.
When obtained as $|U \rbb = | \Phi \rbb \cdot X$, the orthogonality is guaranteed only for the nondegenerate case.

The determination of the principal values and basis can be viewed as finding the set of $N$ closest approximate eigenvalues $\lambda_i$ and eigenstates $|u_i \rangle$ of $\hat{A}$, constrained so that $\spn(U) = \mathcal{V}$. 
As the exact eigenvalue and eigenstate pairs of the projected operator $\hat{A}_{\mathcal{V}}$, the approximation is given so that each residual error $| r_i \rangle = (\hat{A} - \lambda_i) | u_i \rangle$ is orthogonal to $\mathcal{V}$.  

\subsubsection{Operator Variance and Covariance Matrices}
The corresponding operator variance matrices (OVMs) as the promotions of pure state operator variances are formed as matrix analogs of their scalar counterparts.
For a Hermitian operator $\hat{A}$ and basis $\Phi$ of $\mathcal{V}$, the OVM corresponds to $\widetilde{\Sigma}^{(A)}(\Phi) = \widetilde{A}^{(2)}(\Phi) - \widetilde{A}^{2}(\Phi)$, where
$\widetilde{A}^{(2)}$ denotes the OEM for $\hat{A}^2$ and $\widetilde{A}^2 =\widetilde{A} \cdot \widetilde{A}$.
Just as the pure state operator variance for a state $\phi$ can be expressed equivalently as operator expectation of $\hat{A} \hat{P}_{\phi^\perp}\hat{A}$, the OVM can be expressed equivalently as the OEM of $\hat{A}\hat{P}_{\mathcal{V}^\perp}\hat{A}$.
The projector $\hat{P}_{\phi^\perp}$, which projects out the single state $\phi$, is promoted to $\hat{P}_{\mathcal{V}^\perp}$ ,which projects out all the states in $\mathcal{V}$. Operator covariance matrices (OCMs) for two operators $\hat{A}$ and $\hat{B}$ , which we denote as $\widetilde{\Sigma}^{(AB)}(\Phi)$, are formulated similarly.
Hence, we have:
\begin{equation}
    \begin{aligned}
        \widetilde{\Sigma}^{(A)}(\Phi) &= \widetilde{A}^{(2)}( \Phi) -  \widetilde{A}^2(\Phi) =\lbb \widetilde{\Phi}| \hat{A} \hat{P}_{\mathcal{V}^\perp} \hat{A} | \Phi \rbb, \\
        \widetilde{\Sigma}^{(AB)} (\Phi) &= \widetilde{AB}(\Phi)-\widetilde{A}(\Phi)\cdot \widetilde{B}(\Phi) =
        \lbb \widetilde{\Phi}| \hat{A} \hat{P}_{\mathcal{V}^\perp} \hat{B} | \Phi \rbb.
    \end{aligned}
\end{equation}

As the OVM of $\hat{A}$ is the OEM of $\hat{A}\hat{P}_{\mathcal{V}^\perp}\hat{A}$, the principal operator variances $\{ \sigma_k^2 \}$ are not the exact pure state variances of the principal basis $W$.
As for each $w_k$, $\sigma_k^2$ and $\Var_{w_k}[\hat{A}] \equiv \langle \hat{A}^2 \rangle_{w_k} -  \langle \hat{A} \rangle_{w_k}^2$ are given by the expectation values of $\hat{A}\hat{P}_{\mathcal{V}^\perp}\hat{A}$ and $\hat{A}\hat{P}_{w_k^\perp} \hat{A}$ respectively. As $\hat{P}_{\mathcal{V}^\perp}$ projects out $w_k$, as well as the the other states, the two variances are in general not equal but bounded as $0 \leq \sigma_i^2 \leq \Var_{w_i} [\hat{A}]$.
The principal operator covariances are then the expectations of $\hat{A}\hat{P}_{\mathcal{V}^\perp}\hat{B}$ and are similarly bounded in absolute value with the pure state operator covariances. 

The modified variances as expectations of $\hat{A}\hat{P}_{\mathcal{V}^\perp}\hat{A}$ can be obtained individually for any orthogonal basis from the OVM diagonals. Importantly, unique to the principal operator expectation basis $U$, these all equate exactly to the pure state operator variances: 
\begin{equation}
    \widetilde{\Sigma}^{(A)}_{ii}(U) = \Var_{u_i}[\hat{A}].
\end{equation}

From the previous section, each operator image given as $\hat{A}|u_i \rangle = \lambda_i|u_i \rangle + \perp\mathcal{V}$
is already orthogonal to all the other states $u_j$.
Thus, the subsequent projections by either $\hat{P}_{\mathcal{V}^\perp}$ or $\hat{P}_{u_i^\perp} $ are equivalent.

The OVMs can also be understood geometrically as encoding information about how much of the operator image space $\hat{A}\mathcal{V}$ lies outside of $\mathcal{V}$.
This corresponds to the residual space $\mathcal{R} \perp \mathcal{V}$, given so that $\hat{A}\mathcal{V} \subset \mathcal{V} + \mathcal{R}$.
For any basis $\Phi$ of $\mathcal{V}$, a corresponding residual basis of $\mathcal{R}$ can be obtained as $|R\rbb =\hat{P}_{\mathcal{V}^\perp}\hat{A}|\Phi \rbb$.
For orthonormal bases, the OVMs $\widetilde{\Sigma}^{(A)}(\Phi)$ then reduce to their Gram matrix $G(R)$.
The modified operator variances from the diagonals, then, equate to their norms $|\!|r_i|\!|^2$ as a set of individual measures relative to the scalings of $\hat{A}$.
The principal operator variance basis can thus be defined equivalently as the unique orthogonal basis whose residual basis is also orthogonal.
Together, these give respectively the ideal bases for relating $\mathcal{V}$  and $\mathcal{R}$ with one another.

\subsubsection{Monte Carlo Estimations of Operator Matrices}

\begin{figure}[hbtp]
    \centering
    \includegraphics[width=\linewidth]{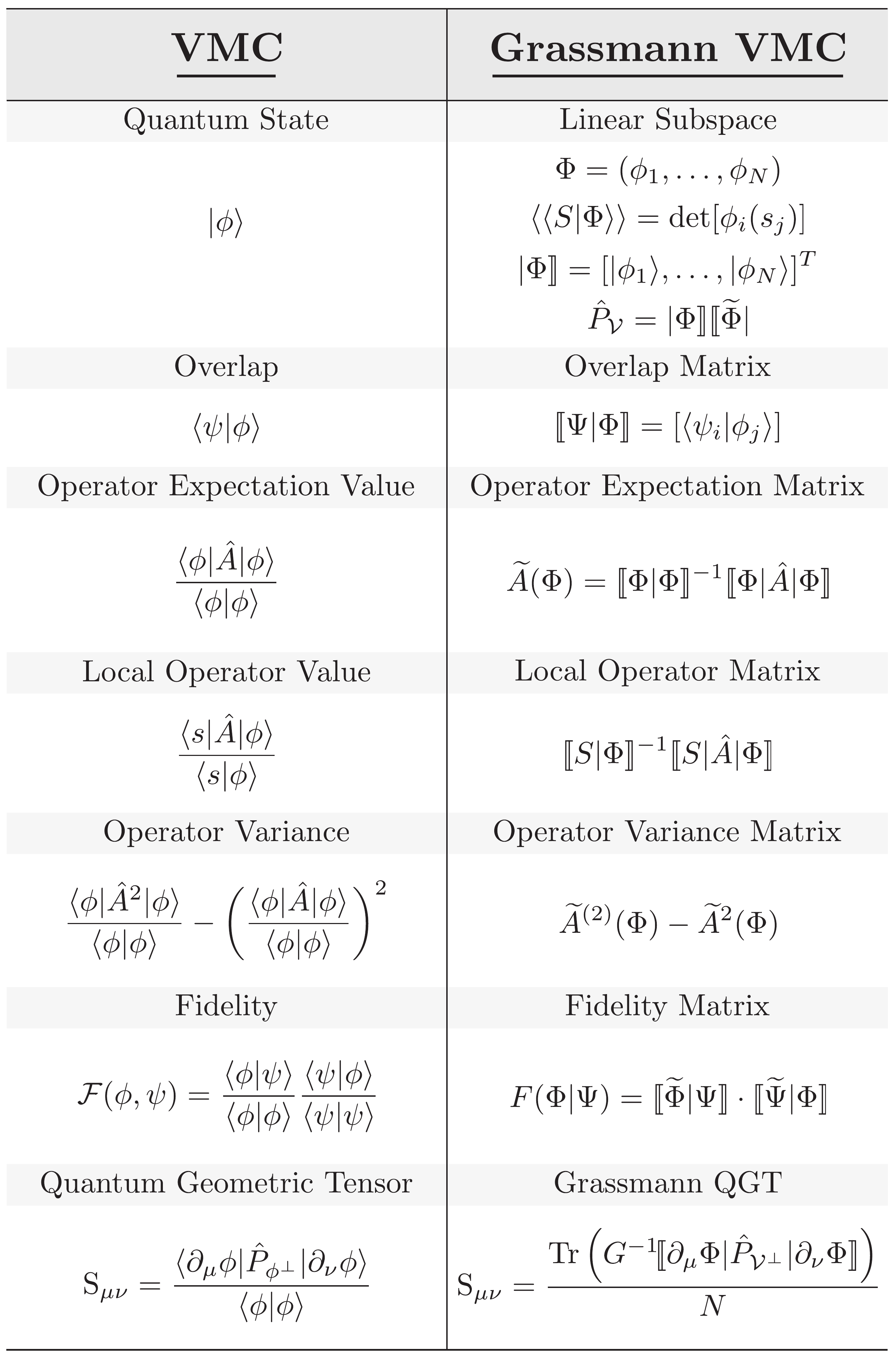}
    \caption{Table of comparison between standard Variational Monte Carlo (VMC) and Grassmann VMC.}
    \label{fig:artistic}
\end{figure}

The standard variational Monte Carlo (VMC) method for estimating operator values for a single quantum state can be extended to the estimations of their matrix promotions using the same Grassmann formulation, which we will refer to as Grassmann variational Monte Carlo (GVMC).
\cref{fig:artistic} shows a table of comparison between VMC and GVMC. 
The fundamentals of GVMC, originally developed by  Pfau et al.~\cite{pfau2024accurate}, are a ``determinant sampling scheme" and a corresponding promotion of the local operator values to matrices.
Together, OEMs can then be estimated directly, like in standard VMC, as the average of the local operator matrices.
In this section, we show that OVMs and OCMs can also be estimated similarly, as an extension of standard VMC.
In a later section, we will show that the same can be done for the estimations of the overlap and the fidelity.

For an orthonormal basis ${ \ket{s} }$ of the Hilbert space, the determinant sampling scheme~\cite{pfau2024accurate} involves drawing configurations in batches, such that a single sample corresponds to a $N$-tuple of basis states $S = (s_1, \ldots, s_N)$.
These are drawn from the distribution of the wedge product representation $P_{\Phi}(S) \propto |\lwa S | \Phi \rwa|^2$ as in standard VMC. 
In the matrix representation, the scalar wavefunction coefficients $\phi(s) =\langle s | \phi \rangle $ become the $N \times N$ overlap matrices $\Phi(S) = \lbb S | \Phi \rbb$ and from~\cref{eq:wedge_inner_det} the wedge product coefficients are their determinants
$\lwa S | \Phi \rwa = \det[\Phi(S) ]$. Conveniently, including all the $N!$ orderings of the individual states, the probability of each $N$-tuple $S$ is then
\begin{equation}
    P_{\Phi}(S) =  \frac{|\det(\Phi(S))|^2}{ N! \det[G(\Phi)] }.
    \label{eq:det_samp}
\end{equation}

The probabilities are invariant under changes of equivalent bases and thus each $S$ is weighted with respect to $\mathcal{V}$.

The determinant sampling allows the expectation value matrix $\widetilde{A}$ to be estimated directly as the Monte Carlo average of the local operator matrix $\widetilde{A}(S)$~\cite{pfau2024accurate}:
\begin{equation}
    \widetilde{A}(\Phi) = \mathbb{E}_{S \sim P_\Phi} [\widetilde{A}(S)].
\end{equation}

For each sampled state tuple $S$, the local operator matrix $\widetilde{A}(S)$ is constructed as a matrix extension of the standard scalar local operator values. 
The operator overlap matrix $A(S) = \lbb S| \hat{A} |\Phi \rbb$ is normalized by the inverse of the overlap matrix $\Phi(S)=\lbb S | \Phi \rbb$ such that:
\begin{equation}
    \widetilde{A}(S) = \Phi^{-1}(S)\cdot A(S).
    \label{eq: OEM VMC}
\end{equation}

Importantly, this bypasses the need for separate estimations of the unnormalized expectation value matrix $A=\lbb \Phi | \hat{A} \Phi \rbb$ and the Gram matrix $G$, as well as its inversion which can be unstable with finite samples.

We now introduce a way of obtaining OVMs and OCMs using the same determinant sampling of local operator matrices.
The matrix extension of this, as dictated by the determinant sampling, becomes the covariance matrix of the trace of one local operator matrix with the individual matrix elements of the other:
\begin{equation}
\begin{aligned}
\widetilde{\Sigma}_{ij}^{(A)}(\Phi) &= \Cov_{S \sim P_\Phi} \left[\Tr (\widetilde{A}(S) ), \widetilde{A}_{ij} (S)  \right], \\
 \widetilde{\Sigma}_{ij}^{(AB)}(\Phi) &= \Cov_{S \sim P_\Phi} \left[\Tr (\widetilde{A}(S) ), \widetilde{B}_{ij} (S)  \right].
\label{eq:cov_mat_samp}
\end{aligned}
\end{equation}

This asymmetric formulation between the local operator matrices is a consequence of the normalization by the Gram matrix inverse on the left. The full covariance between the local operator matrices  $ \Ex_{S \sim P_\Phi}[\widetilde{A}^\dagger(S) \otimes \widetilde{B}(S)] - \Ex_{S \sim P_\Phi}[\widetilde{A}^\dagger(S)] \otimes \Ex_{S^\prime \sim P_\Phi}[\widetilde{B}(S')]$ equates to $G^{-1} \otimes \Sigma^{(AB)}$, but with the relative indices permuted.
Therefore, the contractions that give $\widetilde{\Sigma}^{(AB)} = G^{-1} \cdot \Sigma^{(AB)}$ correspond to taking the traces of $\widetilde{A}^\dagger(S)$.
In an orthogonal basis, the diagonal OVM or OCM matrix elements $\Sigma_{ii}^{(AB)}$ can be obtained alternatively, without the contractions, from the $iiii$ tensor elements of the full covariance. 
This corresponds to the ``variance estimator" described in~\cite{pfau2024accurate}. 
As explained in the previous section, these tensor elements give the modified operator variances and covariances, which only in some cases are the individual pure state operator variances and covariances.
The sampling expressions for the OEMs, OVMs, and OCMs are discussed in detail in~\cref{sec:one_local_matrix,sec:two_local_matrices}.

\subsubsection{Scalar Grassmann Operator Values}

Scalar Grassmann operator values can be defined naturally for each linear subspace $\mathcal{V}$ as the arithmetic means of the principal operator values.
These can then be obtained from the operator matrices as their traces normalized by $1/N$. Without the normalization by $1/N$, these are all equivalently the corresponding pure state operator values on the wedge product state $|\Phi \rwa$, with the operators becoming the directed sums $\hat{A} \rightarrow \hat{A}^{\overline{\oplus}N} =\hat{A} \otimes I \otimes \ldots + I \otimes \hat{A} \otimes \ldots + \ldots$. Additionally, taking the traces of all the GVMC expressions for the operator matrices reduces to the standard VMC expressions for $|\Phi \rwa$, as the traces of the local operator matrices for $\Phi$ equate to the local operator values for $|\Phi \rwa$.
In the density operator representation, the scalar Grassmann operator expectation (with the normalization) is equal to the mixed state operator expectation for $\hat{\rho}_{\mathcal{V}}$, while the scalar Grassmann operator variance and covariance give different definitions than the mixed state ones.

The Grassmann operator expectation value $\langle \hat{A} \rangle_{\mathcal{V}}  = (1/N) \Ex_{S\sim P_\Phi}\left[\Tr[\widetilde{A}(S)]\right]$ is the average of the principal operator expectations, each of which is again the pure state expectation $\langle \hat{A} \rangle_{u_i}$ of the corresponding principal basis state $u_i$. 
As a loss function, $\langle \hat{A} \rangle_{\mathcal{V}} $ corresponds to collectively minimizing all the expectation values $\langle \hat{A} \rangle_{u_i}$ with an orthogonality constraint $\langle u_i | u_j \rangle \propto \delta_{ij}$, which is built in from the Grassmann formulation. 
This is what is done in~\cite{pfau2024accurate} to compute the low-lying excited states of a Hamiltonian $\hat{H}$.  In that case, the minimization of $\langle \hat{A} \rangle_{\mathcal{V}} $ is over a set of $N$ variational wave functions $\Phi$ as an arbitrary basis of $\mathcal{V}$. Then, the linear combinations of the states in $\Phi$, which give the principal basis $U$, form the variational approximations of the excited states.

The Grassmann operator variance $\Var_{\mathcal{V}}[\hat{A}] = (1/N) \Var_{S\sim P_\Phi}\left[\Tr[\widetilde{A}(S)]\right]$ as a loss function can be understood by the equivalent minimal conditions.
\begin{equation}
    \Var_{\mathcal{V}}[\hat{A}] =0 \Leftrightarrow (\hat{A}-\lambda_i)|u_i \rangle = 0 \, \forall \,  i \Leftrightarrow \hat{A} \mathcal{V} \subset \mathcal{V}.
\end{equation}

The average of the principal operator variances is equal to the average of the pure state operator variances of the principal expectation basis states $U$. Thus, minimizing $\Var_{\mathcal{V}}[\hat{A}]$ can be used similarly to collectively minimize each $\Var_{u_i}[\hat{A}]$ with built-in orthogonality constraints. Alternatively, $\Var_{\mathcal{V}}[\hat{A}]$ can be used as a penalty function to enforce the closure requirement $ \hat{A} \mathcal{V} \subset \mathcal{V}$. The mixed state operator variance of $\hat{\rho}_{\mathcal{V}}$ is given as a weighted spread of the eigenvalues of $\hat{A}$. This differs from the Grassmann definition for $\Var_{\mathcal{V}}[\hat{A}]$, which is much less dependent on the particular eigenvalues themselves. An extremal example of the difference is given for $\mathcal{V}$ as the span of the two extremal eigenstates of $\hat{A}$. The Grassmann variance is minimally zero as the principal expectation states are exact eigenstates, while the mixed state operator variance is maximally $\left(\frac{1}{2}(\lambda_{\text{max}} -\lambda_{\text{min}}) \right)^2$, with $\lambda_{\text{min}}$ and $\lambda_{\text{max}}$ minimal and maximal eigenvalues respectively. 

\subsubsection{Wedge-Product States as Free Fermion States}
We remark that a linear subspace in the wedge-product representation can be interpreted as a state of $N$ free fermions where single particle states are replaced by many-body wave functions.
Its corresponding un-normalized amplitudes are given by:
\begin{equation}
    \langle \langle S | \Phi \rangle \rangle = \det \lbb S | \Phi \rbb.
\end{equation}

One can then define an extended Hamiltonian in $\mathcal{H}^{\overline{\otimes} N}$ as $\mathbb{H} = \hat{H} \otimes \hat{I} \otimes \ldots \otimes \hat{I} + \hat{I} \otimes \hat{H} \otimes \ldots \otimes \hat{I} + \ldots + \hat{I} \otimes \hat{I} \otimes \ldots \otimes \hat{H}$, where $\hat{H}$ is the physical Hamiltonian of interest.
Since $\mathbb{H}$ is the sum of $N$ non-interacting terms, its eigenstates are tensor product of the eigenstates of $\hat{H}$.
In particular, the ground state of $\mathbb{H}$ with fermionic symmetry is the Slater determinant of the first $N$ excited states of $\hat{H}$.

\subsection{Grassmann Geometry}

\label{secsec:Local Metrics and Gradients }
\subsubsection{Fidelity Matrices}

The fidelity between two quantum states $\mathcal{F}(\phi, \psi)=\bra{\phi}\ket{\psi}\bra{\psi}\ket{\phi} / \bra{\phi}\ket{\phi}\bra{\psi}\ket{\psi}$
is extended to the case of two bases $\Phi$ of $\mathcal{V}$ and $\Psi$ of $\mathcal{V}'$ by promoting scalar overlaps to overlap matrices, following the same procedure used in the previous generalizations.
The only difference is in the last step, when selecting the matrix multiplication orderings which transform equivariantly, as now the similarity transformations for both  $|\Phi \rbb \rightarrow |\Phi \rbb \cdot X$ and $|\Psi \rbb \rightarrow |\Psi \rbb \cdot Y$ have to be taken into account. 
The result of the matrix promotion is two matrices with equivalent eigenvalues, which we denote as
$F(\Phi \, | \, \Psi)$ and $F(\Psi \, | \, \Phi)$. 
The total equivariance is then given individually so that $F(\Phi | \Psi)$ ($F(\Psi | \Phi)$) transforms equivariantly with respect to $\Phi$ ($\Psi$) and invariantly with respect to $\Psi$ ($\Phi$). 

The orderings which give these two respective combinations of equivariance and invariance are the two orderings of the normalized overlap matrices $\lbb \widetilde{\Phi}|\Psi \rbb$ and $\lbb \widetilde{\Psi}|\Phi \rbb$. 
As the orderings can be related by a similarity transformation, their eigenvalues are the same. Analogously to their scalar counterparts, the two fidelity matrices are equivalently the two OEMs of the projectors $\hat{P}_{\mathcal{V}'}$ and $\hat{P}_{\mathcal{V}}$ for $\Phi$ and $\Psi$ respectively.
\begin{equation}
    \begin{aligned}
        F(\Phi | \Psi)  &= \lbb \widetilde{\Phi}| \Psi \rbb \cdot  \lbb \widetilde{\Psi}| \Phi \rbb =  \widetilde{P}_{\mathcal{V}'}(\Phi),   & \mathcal{V'} = \spn(\Psi) \\
        F(\Psi | \Phi)  &=    \lbb \widetilde{\Psi}| \Phi \rbb \cdot \lbb \widetilde{\Phi}| \Psi \rbb = \widetilde{P}_{\mathcal{V}}(\Psi),   &\mathcal{V} = \spn(\Phi).
    \end{aligned} 
    \label{eq:fid}
\end{equation}

From their shared eigenvalues, the principal fidelities $\{ \lambda_i^2 \}$ are defined equivalently from either fidelity matrix.
From the combination of equivariance and invariance,  principal bases $U$ of $\mathcal{V}$ and $W$ of $\mathcal{V}'$ are defined so that they each diagonalize their respective fidelity matrices as $F(U | \Psi) = \Lambda^2$ and $F(W | \Phi) = \Lambda^2$.
The principal fidelities and bases are then obtained from the eigendecompositions of the fidelity matrices $ F(\Phi|\Psi) = X \cdot \Lambda^2 \cdot X^{-1}$ and $ F(\Psi|\Phi) = Y \cdot \Lambda^2 \cdot Y^{-1}$ with  $|U\rbb = |\Phi \rbb \cdot X$ and $|W \rbb = |\Psi \rbb \cdot Y$.

The principal fidelities are the exact pure state fidelities of the principal bases pairs $ \lambda_i^2 =\mathcal{F}(u_i,w_i)$. 
The corresponding Fubini-Study angles $ \alpha_i = \cos^{-1}(\lambda_i) $ give the set of Grassmann principal angles, also known as the angles between flats.
These are the fundamental set of angles between $\mathcal{V}$ and $\mathcal{V}'$ as hyperplanes. The principal bases $U$ and $W$, which give the corresponding direction pairs of these angles, form the ideal orthogonal bases for $\mathcal{V}$ and $\mathcal{V}'$ respectively for relating the two spaces to each other. Algebraically, this is formulated as $U$ and $W$ are the orthogonal bases whose mutual overlaps are $\langle u_i | w_j \rangle \propto \delta_{ij}$, which is ensured by the respective diagonalization of the fidelity matrices. 
Scaled as orthonormal bases with the right chosen relative phases, both overlap matrices can be expressed as $\lbb U | W \rbb = \lbb W | U \rbb = \Lambda $.
The states of $W$ can then be expressed in terms of $U$ as $|w_i \rangle = \cos(\alpha_i)|u_i \rangle + \sin(\alpha_i)|r_i \rangle $, where the states $|r_i \rangle \propto \hat{P}_{\mathcal{V}^\perp} |w_i \rangle$ are normalized as an orthonormal basis for the residual space $\mathcal{R} \perp \mathcal{V}$ which gives $\mathcal{V}' \subset \mathcal{V} +\mathcal{R}$. 

\subsubsection{ Monte Carlo Estimations of Overlap and Fidelity Matrices}

Here we introduce a VMC method for estimating normalized overlap and fidelity matrices. Like for operator value matrices, we use the same determinant sampling scheme~\cref{eq:det_samp} and formulate the sampling expressions as matrix extension of the standard VMC expressions. 

For single state pairs $\phi$ and $\psi$, sampled in terms of $\phi$, the local values are given by the coefficient ratios $r(s) = \psi(s) /\phi(s) $. 
Now for bases $\Phi$ and $\Psi$, sampled in terms of $\Phi$, the local matrices are $R(S) = \Phi^{-1}(S) \cdot \Psi(S)$. 
The normalized overlap matrix $\lbb \widetilde{\Phi}| \Psi \rbb$ is given analogously to $\langle \phi | \psi \rangle / \langle \phi | \phi \rangle $, as the average of the local matrices $R(S) $. Then $\lbb \widetilde{\Psi}| \Phi \rbb$ is given as the average of $R^{-1 }(S) $ either by sampling directly in terms of $\Psi$ or in terms of $\Phi$ and importance weighting. 
Therefore, we can write:  
\begin{equation}
    \begin{aligned}
        \lbb \widetilde{\Phi}| \Psi \rbb &= \Ex_{S \sim P_{\Phi}} [ \, R(S) \,  ], \hspace{0.5cm} \lbb \widetilde{\Psi}| \Phi \rbb = \Ex_{S \sim P_{\Psi}} [ \, R^{-1}(S) \,  ].
    \end{aligned}
    \label{eq: Overlap VMC}
\end{equation}

The importance sampling expression becomes: 
\begin{equation}
    \lbb \widetilde{\Psi}| \Phi \rbb =  \frac{\Ex_{S \sim P_{\Phi}} \left[ \, |\det(R(S))|^2 R^{-1}(S) \,  \right]}{ \Ex_{S^\prime \sim P_{\Phi}} \left[ |\det(R(S^\prime))|^2 \right]}, 
\end{equation}
and equivalently for the other matrix. 

The two fidelity matrices can then be estimated by multiplying the respective sampling expressions for the overlap matrices. This is the matrix analog of writing the standard VMC expression for the fidelity $\mathcal{F} =|\Ex[r(s)]|^2 / \Ex[|r(s')|^2]$ as the multiplication of $\langle \phi | \psi \rangle / \langle \phi | \phi \rangle$ as $\Ex[r(s)] $ by $\langle \psi | \phi \rangle / \langle \psi | \psi \rangle$ using importance sampling as $\Ex[|r(s)|^{2 \, } r^{-1}](s) / \Ex[|r(s')|^{2}]$. 
Alternatively, a different sampling expression for the fidelity matrix $F(\Phi|\Psi)$ can be obtained that does not require importance sampling.  
This is formulated as the matrix analog of the standard VMC formulation, rearranged as an estimator of $\mathcal{F}^{-1} -1$.
For the scalar case, this becomes $\Var[\, r(s) / \Ex[r(s')]\,]$, and for the matrix case this becomes a wholly different sampling expression for $F^{-1}(\Phi|\Psi) -1$. Firstly, the normalization by the averages is given as $\Gamma(S) = R(S)  \cdot \left( \Ex[ R(S') ] \right)^{-1}$. Then the ``variance" of the normalized local matrices $\Gamma(S)$ is given identically as for OVMs  and local operator matrices~\cref{eq:cov_mat_samp}, where the covariance matrix of the local matrices' total traces with their individual matrix elements is the general local matrix analog of the variance of scalar local values:
\begin{equation}
    \begin{aligned}
         F^{-1}(\Phi|\Psi) -1  = \Cov_{S \sim P_{\Phi}}\left[ \, \Tr( \Gamma(S) ) \, , \,  \Gamma(S) \, \right].
    \end{aligned}
\end{equation}

\subsubsection{Metric Angle and Scalar Fidelity }
Grassmann metrics can be defined analogously to the Fubini-Study metric for rays as sign-less angles between linear subspaces. The metric angles are constructed naturally as some averages of the individual principal angles, as Fubini-Study metric angle themselves. 
However, as there are many choices of average, there is no singular definition of a Grassmann metric angle and many definitions can be found in the math literature.
Deriving a metric angle $\alpha$, or equivalently a scalar fidelity as $\cos^2(\alpha)$, from the various Grassmann representations will also give different definitions depending on the chosen representation.

All the relevant definitions can be expressed as different generalized/power means of $\cos(\alpha_i)$. Fortunately, as we will show in the next section, these will give the same metric tensor. 
We denote the metric angle for the power mean $p$-mean as $\alpha^{(p)}(\mathcal{V},\mathcal{V}')$ so that 
\begin{equation}
    \cos \small( \alpha^{(p)} \small) =  \left( \frac{1}{N} \sum_{i=1}^{N} \cos^p ( \alpha_i) \right)^{1/p}.
\end{equation}

The corresponding scalar $p$-fidelity given as $\cos^2 (\alpha^{(p)})$, is equivalent to the $p/2$-mean of the principal fidelities. 
This can be expressed in terms of fidelity matrices for any two bases $\Phi$ of $\mathcal{V}$ and $\Psi$ of $\mathcal{V}'$ as
\begin{equation}
     \mathcal{F}^{(p)}(\Phi,\Psi) = \left( \frac{1}{N}  \Tr \left[ F^{\frac{p}{2}}(\Phi|\Psi) \right] \right)^{\frac{2}{p}}.
    \label{eq:p_fid_def}
 \end{equation}

The power $p$ parameterizes the relative weighting of the principal angles/fidelities such that as $p$ increases more weight is given to the smaller angles/larger fidelities. 
$p=1$ is the ``Goldilocks" weighting, which gives the ideal gradient as individual Fubini-Study gradients concatenated in the principal bases.
The power means $p=0,1,2$ can each be related to the Grassmann representations. 
In particular, we have:
\begin{itemize}
\item \textbf{$0$-Fidelity (Geometric Mean): Wedge-Product Representation}. 
The pure state fidelity between the two wedge product states $|\Phi \rwa$ and $|\Psi \rwa$ is given by the determinant of either fidelity matrix and $\mathcal{F}^{(0)}(\Phi,\Psi)$ is the $N$-th root of the determinant.

\item \textbf{$1$-Fidelity (Arithmetic Mean): Density Operator Representation}.
The mixed state fidelity between the density operators 
$\left(  \Tr  \sqrt{ \hat{\rho}_{\mathcal{V}} \cdot \hat{\rho}_{\mathcal{V}'}   }  \right)^2$
is equal to $\mathcal{F}^{(1)}(\Phi,\Psi)$.

\item \textbf{$2$-Fidelity (Root Mean Square): Projector Representation}.
The Frobenius inner product  between the orthogonal projectors $ \Tr \small( \hat{P}_{\mathcal{V}} \hat{P}_{\mathcal{V'}} \small)$ is proportional by $1/N$ to $\mathcal{F}^{(2)}(\Phi,\Psi)$ .

\end{itemize}

\subsubsection{Metric Tensor and Tangent Spaces}
Though there are multiple definitions of Grassmann metric angles, there is a singular standard definition of the Grassmann metric tensor which is ubiquitous in the literature.
This is because the infinitesimal form metric angles $\alpha^{(p)}$, from which the corresponding metric tensor is derived, reduce to the same values for all powers $p$.
As a result, there is no ambiguity in defining tangent space quantities such as gradients and projections.

Formally, a metric tensor is defined from an inner product between pairs of tangent space vectors, so that the corresponding vector norm relates to the infinitesimal form of the metric.
Grassmann tangent space vectors can be expressed (not uniquely) in a basis representation  $\Phi$ as $N$-tuples $\delta \Phi$ of individual basis state perturbations.
The induced norm $|\!|\delta \Phi |\!|_\Phi =\sqrt{\langle \delta \Phi ,\delta \Phi \rangle_\Phi} $ on these perturbations is given so that $\left( \alpha^{(p)}(\Phi,\Phi+\delta\Phi)  \right)^2 = |\!|\delta \Phi |\!|_\Phi^2 + \mathcal{O}(|\!|\delta \Phi |\!|_\Phi^4)$ and is again independent of the power $p$. 
The general form of the metric tensor as inner product between any pair of basis perturbations can be expressed compactly in the matrix representation as  
\begin{equation}
    \langle \delta \Phi ,\delta \Phi'  \rangle_\Phi =\frac{1}{N} \Tr \left( G^{-1}(\Phi) \cdot \lbb \delta \Phi |  \hat{P}_{\mathcal{V}^\perp} | \delta \Phi' \rbb  \right).
\label{eq:metric_tensor}
\end{equation}

Functionally, the Grassmann metric tensor is a matrix promotion of the Fubini-Study metric tensor and can be derived equivalently using the previous techniques for operator values and fidelity.
Similarly, it can also be derived as the Fubini-Study metric tensor in the wedge product representation, but again without the proportionality factor of $1/N$.
This factor is also often not included in definitions found in the math literature, but we opt to keep it as it normalizes the inner products to be independent of the dimension and aligns with previous definitions for scalar operator values.

The abstract changes of directions of linear subspaces, which form the Grassmann tangent space $\mathcal{T}_{\mathcal{V}}\mathbf{Gr}_N(\mathcal{H})$, can be expressed in the basis representation as perturbations $\delta \Phi$ to the given basis $\Phi$ of $\mathcal{V}$.
However, this representation is not unique as different basis perturbations can result in the same linear subspace to first order.
In order to define a bijective representation, the components of the basis perturbation that correspond to a change of equivalent basis as $|\delta \Phi \rbb = |\Phi \rbb \cdot \delta X $ for any $ \delta X \in GL_N (\mathbb{C})$ must be removed.
This equivalently enforces that each individual basis state perturbation $|\delta \phi_i \rangle$ is orthogonal to $\mathcal{V}$ and thus $\mathcal{T}_{\mathcal{V}}\mathbf{Gr}_N(\mathcal{H}) \simeq (\mathcal{V}^{\perp})^N $.
In the matrix representation, the tangent space ket matrices formed so that all their column vectors are in $\mathcal{V}^{\perp}$ are given collectively as $\mathcal{T}_{\mathcal{V}}\mathbf{Gr}_N(\mathcal{H}) \simeq \{ |\delta \Phi \rbb \, | \, \lbb \widetilde{\Phi}| \delta \Phi \rbb = \boldsymbol{0}\}$, where $\boldsymbol{0}$ indicates the matrix with zero entries.
This is a higher-dimensional analog of the usual tangent space representation of the projective Hilbert space $\mathcal{P}(\mathcal{H})$, where the tangent space of a ray, represented by a given state $\phi$, is expressed bijectively as $\mathcal{H}_{\phi^\perp}$~\cite{hackl2020geometry}.
In the wedge product representation, the tangent space vectors are given bijectively as $|\delta \phi_{1} \rangle \wedge | \phi_{2} \rangle \wedge \ldots \wedge |\phi_{N} \rangle + | \phi_{1} \rangle \wedge | \delta \phi_{2} \rangle \wedge \ldots \wedge |\phi_{N} \rangle + \ldots + |\phi_{1} \rangle \wedge | \phi_{2} \rangle \wedge \ldots \wedge | \delta \phi_{N} \rangle$.

\section{Variational Grassmanians}
In this section, we develop the framework for the variational representation of Grassmannians, extending the optimization of a single variational wave function to the simultaneous optimization of multiple ones. 

\subsection{Constructing Variational Linear Subspaces}

We will assume a variational construction of linear subspaces by the span of $N$ individual variational wave functions (VWFs) with an assumed linear independence.
The variational linear subspace $\mathcal{V}(\theta) = \spn(\Phi(\theta))$ is then parametrized by the combined $M$ VWF weights $\theta=(\theta^{1},\ldots,\theta^{M} )$, which are assumed to be real and differentiable or complex and holomorphic. We will not assume whether each individual weight $\theta^\mu$ is unique to a specific VWF or shared between multiple VWFs, as there are advantages and disadvantages to both constructions.
We consider the VWFs to be represented by neural quantum states (NQS) wave functions~\cite{carleo2017solving}. 
A NQS basis given by $N$ networks with identical architectures and separate sets of $M / N$ individual weights has ideal variational properties and gives block diagonal Jacobians.
However, as the calculation of a single overlap matrix $\Phi(S) = \lbb S | \Phi \rbb$ requires $N^2$ forward passes, then the computational cost of a VMC calculation with deep neural networks (DNNs) ansätze can be prohibitive for high-dimensional subspace. 
The overhead can be significantly reduced by sharing weights among the models. 
For instance, we can consider a NQS basis given collectively by a single shared DNN base — acting as a common backflow transformation~\cite{diluo2019} — paired with $N$ individual shallow neural network (SNN) heads. 
In this case, evaluating $\Phi(S)$ requires only $N$ DNN and $N^2$ SNN forward passes, substantially lowering the total cost.
This is precisely the architecture we decided to adopt in our simulations (see~\cref{sec:results}).

\subsection{The Variational Manifold}
    
Most standard VMC schemes for a single VWF $\phi(\theta)$, such as the Stochastic Reconfiguration method~\cite{sorella1998green}, can be formulated in terms of the differential geometry of the variational manifold $\mathcal{M}$~\cite{hackl2020geometry}, where $\mathcal{M} \subset \mathcal{P}(\mathcal{H})$ formed as the sub manifold of the projective Hilbert space given by all the corresponding rays containing each possible VWF $\phi(\theta)$.
The VWF weights $\theta$ are then viewed as a working set of coordinates for $\mathcal{M}$.
Variational problems are thus formulated in purely geometric terms, as the search for an optimal ray constrained to the variational manifold $\mathcal{M}$.
The VWF optimal weights are then given just as the coordinates of the ray solution.
Using this geometric formulation, standard single VWF VMC methods can be generalized to multiple VWFs by fundamentally replacing projective geometry with Grassmann geometry. 
The Grassmann variational manifold  $\mathcal{M} \subset \mathbf{Gr}_N(\mathcal{H}) $ is given analogously as the set of all the possible variational linear subspaces.
Constructed from a VWF bases as  $\mathcal{V}(\theta) = \spn( \Phi(\theta) )$, the collective VWF weights $\theta$ parameterize $\mathcal{M}$ as a set of coordinates. 

\subsection{Variational Tangent Spaces}
The variational tangent space $\mathcal{T}_{
\mathcal{V}(\theta)} \mathcal{M} $ at each $\mathcal{V}(\theta)$ is given by projecting the larger Grassmann tangent space $\mathcal{T}_{\mathcal{V}(\theta)}  \mathbf{Gr}_N(\mathcal{H})$ to only directions which stay in the variational manifold $\mathcal{M}$.
These directions, as first order changes to $\mathcal{V}(\theta)$, are all given uniquely by a corresponding perturbation in the VWF weights $\delta \theta$ and the resulting first order changes to the VWF basis $ \phi_i(\theta +\delta \theta) - \phi_i(\theta) = \delta\theta^\mu \partial_\mu \phi_i(\theta) + \mathcal{O}(\delta \theta^2)$.
The variational tangent vectors are then given by the spans of the coordinate basis vectors formed individually by the partial derivatives for each $\theta^\mu$.  However, as the Grassmann tangent space is bijectively represented by only basis perturbations which are orthogonal to $\mathcal{V}(\theta)$, the components of the partial derivatives parallel to $\mathcal{V}(\theta)$ must be projected out.
In the matrix representation, the effective coordinate basis is then given as
 \begin{equation}
    \mathcal{T}_{
\mathcal{V}(\theta)} \mathcal{M} \simeq \spn \left \{ \, |\Theta_\mu(\theta) \rbb =\hat{P}_{\mathcal{V}^\perp} | \partial_\mu \Phi(\theta) \rbb \, \right \}_\mu.
\end{equation}

This coordinate basis formulation is again just the matrix analog of the standard single state formulation~\cite{hackl2020geometry}, where the effective projective coordinate basis vectors are given by the states $\hat{P}_{\phi^\perp}| \partial_\mu \phi \rangle$.
In the wedge product representation, the effective Grassmann coordinate basis vectors is then $\hat{P}_{\wedge_i \phi_i ^\perp} | \partial_\mu \Phi \rwa = \hat{P}_{\mathcal{V}^\perp}|\partial_\mu \phi_{1} \rangle \wedge | \phi_{2} \rangle \wedge \ldots \wedge |\phi_{N} \rangle + |\phi_{1} \rangle \wedge \hat{P}_{\mathcal{V}^\perp} |\partial_\mu \phi_{2} \rangle \wedge \ldots \wedge |\phi_{N} \rangle + \dots + |\phi_{1} \rangle \wedge |\phi_{2} \rangle \wedge \ldots \wedge \hat{P}_{\mathcal{V}^\perp}|\partial_\mu \phi_{N} \rangle$. 

\subsection{Quantum Geometric Tensor}
The most straightforward geometric definition of the Quantum Geometric Tensor (QGT)~\cite{stokes2020quantum,park2020geometry,hackl2020geometry} is as the induced metric tensor on the variational manifold.
While for the single state it is induced by the Fubini-Study metric, the Grassmann QGT is then naturally induced from the Grassmann metric~\cref{eq:metric_tensor}.
The QGT matrix elements $\QGT_{\mu \nu}$ are given by the inner products of the corresponding coordinate basis pairs $\Theta_\mu$ and $\Theta_\nu$.
This allows the inner product~\cref{eq:metric_tensor}, restricted to pairs of variational tangent space vectors $|U \rbb = U^\mu |\Theta_\mu \rbb $ and $|W \rbb = W^\nu |\Theta_\nu \rbb $ to be expressed as $\QGT_{\mu \nu} \overline{U}^\mu W^\nu$. 
For complex weights and holomorphic VWFs, the Grassmann QGT is given as the Hermitian positive definite matrix
\begin{equation}
\label{eq:QGT_Mat}
    \QGT_{\mu \nu} = \frac{1}{N} \Tr \left(  G^{-1} \lbb \partial_\mu \Phi | \hat{P}_{\mathcal{V}^\perp} |\partial_\nu \Phi \rbb   \right).
\end{equation}

For real parameters, only the real part of~\cref{eq:QGT_Mat} must be taken, which gives a symmetric positive definite matrix.

Another interpretation of the QGT, which is practically important for VMC estimations, is as the covariance matrix of the partial derivatives.
The Grassmann QGT maintains this relationship in terms of the previously defined Grassmann operator covariance, so that $\QGT_{\mu \nu}(\theta) = \Cov_{\mathcal{V}(\theta)}[\partial_\mu,\partial_\nu] $.
Defining $\widetilde{D}_\mu(S) = \Phi^{-1}(S) \cdot \partial_\mu\Phi(S)$ and using the same sampling formulation for operator covariances, the QGT matrix elements can be calculated as
\begin{equation}
    \QGT_{\mu \nu} = \frac{1}{N} \Cov_{S \sim P_{\Phi}} \left[ \,  \Tr (\widetilde{D}_\mu(S)) \, ,  \, \Tr (\widetilde{D}_\nu(S)) \, \right].
\end{equation}

This sampling expression is equivalent to the standard VMC one in the wedge product representation, as $\Tr [\widetilde{D}_\mu(S)]=\partial_\mu \log( \lwa S | \Phi \rwa )$.
This follows from the overlap-determinant relation $\lwa S | \Phi \rwa \propto \det[ \Phi(S)]$ and from the formula for the derivatives of the determinant $\partial_\mu \log( \det[\Phi(S)] ) = \Tr[ \Phi^{-1}(S)\cdot \partial_\mu \Phi(S) ]$.

\begin{figure*}[hbtp]
    \centering
    \includegraphics[width=\linewidth]{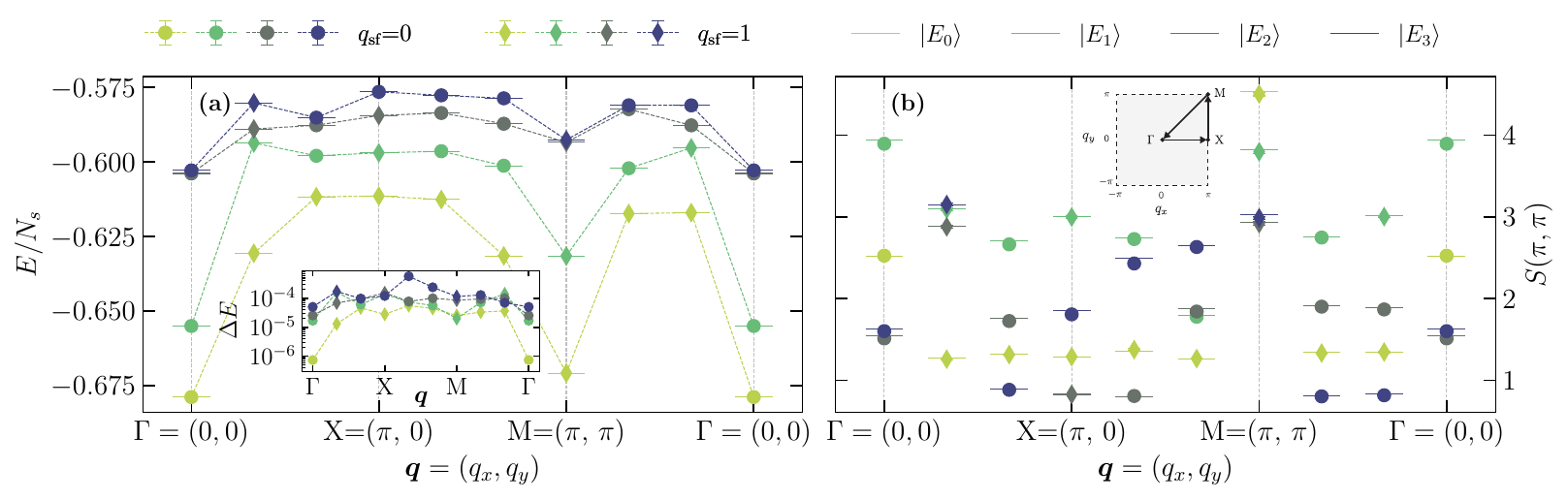}
    \caption{\textbf{(a)} Energy density $E/N_s$ and \textbf{(b)} spin structure factor $S(\pi, \pi)$ for the first 4 excited states of the Heisenberg model on the $6 \times 6$ lattice in different momentum $\boldsymbol{q}=(q_x, q_y)$ symmetry sectors. 
    The markers correspond to the variational results, while the horizontal lines indicate the values from exact diagonalization (ED). 
    The inset displays the energy relative errors $\Delta E = |E - E_{\text{ED}}| / |E_{\text{ED}}|$.
    Different markers stand for different spin-flip symmetry sectors identified by the quantum number $q_{\text{sf}}=0, 1$.
    }
    \label{fig:6x6}
\end{figure*}

\subsection{Optimization of Multiple Wave Functions and Stochastic Reconfiguration}
The extension of the optimization through Stochastic Reconfiguration (SR)~\cite{sorella1998green} to linear subspaces is straightforward using this geometric formalism.
Loss functions $\mathcal{L}$, originally defined over individual wave functions, are now formulated over bases, $\Phi \rightarrow \mathcal{L}(\Phi)$, and must remain invariant under changes of basis to be well-defined on linear subspaces.
For VWF bases, the composite mapping gives the loss as a function over VWF weights: $\theta  \rightarrow \mathcal{L}(\Phi(\theta))$.
The SR weight update $\Delta \theta$ in the Grassmann setting is given identically as for standard SR as the solution of the following equations:
\begin{equation}
    \QGT_{\mu \nu }(\theta) \,  \Delta \theta^\nu = -\eta \frac{\partial \mathcal{L} }{\partial \overline{\theta}^{\mu} }(\Phi(\theta)), 
    \label{eq: SR update}
\end{equation}
where $\eta$ denotes the step size or learning rate.
This expression corresponds to natural gradient descent, where the QGT plays the role of the metric tensor, and the update follows the gradient adapted to the geometry of the variational manifold.
In terms of tangent vectors in the matrix representation, the gradient of $\mathcal{L}$ defined over the variational manifold $\mathcal{M}$ is $|\mathcal{L}_{\mathcal{M}} / \delta \Phi \rbb = (\QGT^{-1})^{\mu \nu} \partial_\nu \mathcal{L}  \, | \Theta_\mu \rbb $.

More generally, SR can also be viewed as approximating some target evolution while being constrained to the variational manifold.
In the continuous-time setting, the approximate velocity at each time is given by projecting the target velocity onto the variational manifold.
For loss minimization, the target evolution is the unconstrained gradient descent direction $| \delta \mathcal{L} / \delta \Phi \rbb$ defined over the full Grassmann manifold $\mathbf{Gr}_N(\mathcal{H})$. Projecting this onto $\mathcal{M}$ yields the gradient restricted to the variational submanifold.
The projected Grassmann tangent space where the variational velocity is given by 
\begin{equation}
    \mathcal{P}_{\mathcal{T}_{\mathcal{V}(\theta)}\mathcal{M} }( |W \rbb ) = | \Theta_\mu \rbb (\QGT^{-1} (\theta) )^{\mu \nu} \, \langle \Theta_\nu,  W \,  \rangle_{\mathcal{V}(\theta)}.
    \label{eq: tangent projection}
\end{equation}

\section{Results}
\label{sec:results}
To test the efficacy of the Grassmann VMC framework formalized in the previous sections, we apply it to find the excited states of a quantum spin Hamiltonian, in the same fashion as done in~\cite{pfau2024accurate} for continuous-space systems.
Let us consider an Hilbert space $\mathcal{H}$ of a system of $N_s$ spins-$\frac{1}{2}$ with basis states $\{\ket*{s}\} = \{\ket*{s^1, \ldots, s^{N_s}}\}$, where $s^i = \pm 1$ for $i = 1, \ldots, N_s$. 
We focus on a two-dimensional spin lattice such that $N_s = L \times L$, where $L$ is the linear dimension of the lattice.
The basis wave functions $\phi_i$ are parametrized as neural wave functions with the following architecture:
\begin{equation}
    \phi_{i}(s) = f_{\theta_i}(\text{CNN}_{\Theta}(s)), 
\end{equation}
where $\theta_i$ and $\Theta$ are real parameters. 
The internal CNN corresponds to a real-valued convolutional neural network inspired by the ConvNeXt architecture~\cite{liu2022convnet,woo2023convnext}, already employed in the context of ground state search~\cite{sinibaldi2025non,nutakki2025design}.
Given an input spin configuration $s \in [-1, 1]^{L \times L}$ the network begins with a convolutional layer using a filter of size $K \times K$, unit stride, and periodic boundary conditions, producing a feature map $y \in \mathbb{R}^{L \times L \times C}$, where $C$ denotes the number of channels.
The network then processes this feature map through $G$ stages.
Each stage first applies a dense layer projecting the input to $w \in \mathbb{R}^{L \times L \times d}$, followed by a sequence of $B$ residual blocks.
For each block, denoting its input as $x_{\text{in}}$, the following operations are performed in order:

\begin{itemize}
    \item depth-wise convolution with $K_{dw} \times K_{dw}$ filter $x_{\text{in}} \to x^{(1)} \in \mathbb{R}^{L \times L \times d}$,  
    \item layer normalization~\cite{ba2016layer}, 
    \item dense layer $x^{(1)} \to x^{(2)} \in \mathbb{R}^{L \times L \times (f \cdot d)}$ realizing the inverted bottleneck with expansion ratio $f$, 
    \item GELU activation function, 
    \item global response normalization~\cite{woo2023convnext}
    \item dense layer $x^{(2)} \to x^{(3)} \in \mathbb{R}^{L \times L \times d}$,
    \item rescaling $x^{(4)} = \gamma \cdot x^{(3)}$ where $\gamma$ is a learnable parameter, 
    \item residual connection $x_{\text{out}} = x_{\text{in}} + x^{(4)}$.
\end{itemize}

The final output of all the stages $z \in \mathbb{R}^{L \times L \times d}$ is then subjected to a last layer normalization, flattened and given as input to the $f_{\theta_i}$ which correspond to complex-valued Restricted Boltzmann Machines (RBMs)~\cite{carleo2017solving}. 
The CNN acts as a common backflow transformation of the spin coordinates which is shared by all the ansätze, while the RBMs are specific for each variational wave function. 
This approach is in line with the \emph{representational learning} paradigm~\cite{bengio2013representation}  already
employed in previous works on ground state~\cite{viteritti2023transformer_2d,denis2024accurate}.
The CNN model is equivariant with respect to translational and spin-flip symmetries of the input, while the RBMs implement momentum and spin projection, making the final wave functions invariant under these symmetries. 
The ansätze are optimized using the SR method for Grassmannians formalized in the previous section.
Similarly to the single-state case, we also leverage the Woodbury matrix identity to speed up SR updates in the limit of number of samples much smaller than the number of variational parameters~\cite{chen2024empowering,rende2024simple}. 
The sampling is restricted to the sector with zero total $z$-magnetization.
The Marshall sign-rule~\cite{marshall1955antiferromagnetism} is applied to the outputs of the neural networks. 

\begin{figure}[hbtp]
    \centering
    \includegraphics[width=1.0\linewidth]{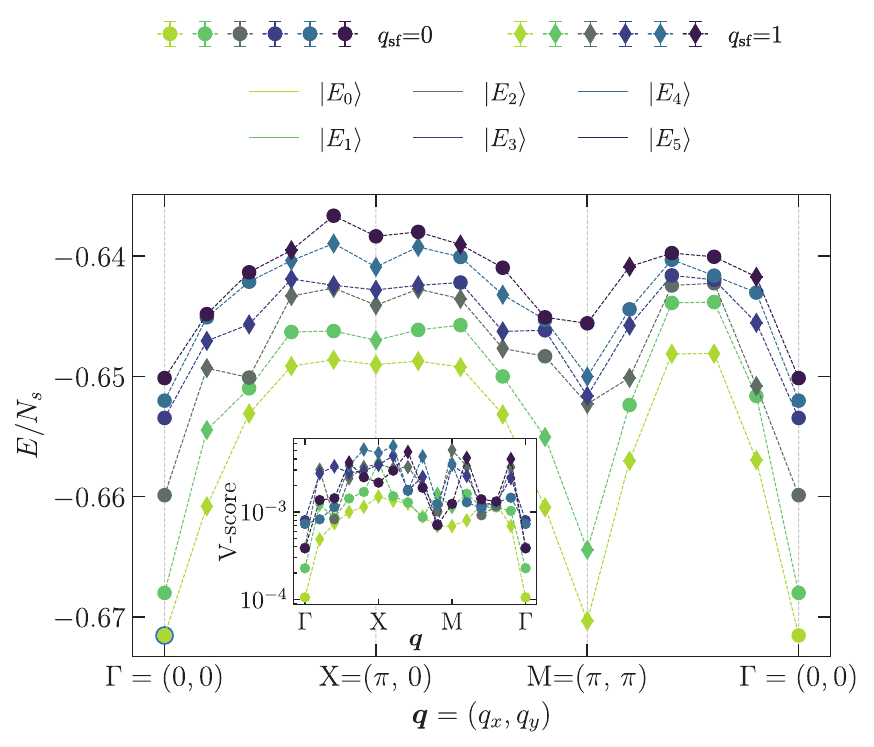}
    \caption{Energy density $E/N_s$ for the first 6 excited states of the Heisenberg model on the $10 \times 10$ lattice in different momentum $\boldsymbol{q}=(q_x, q_y)$ symmetry sectors. 
    The inset displays the corresponding V-scores as measures of the variational accuracy.
    Different markers stand for different spin-flip symmetry sectors identified by the quantum number $q_{\text{sf}}=0, 1$.
    The blue circle indicates the ground state energy from Quantum Monte Carlo calculation~\cite{sandvik1997finite}.
    }
    \label{fig:10x10}
\end{figure}

\begin{figure}[hbtp]
    \centering
    \includegraphics[width=1.0\linewidth]{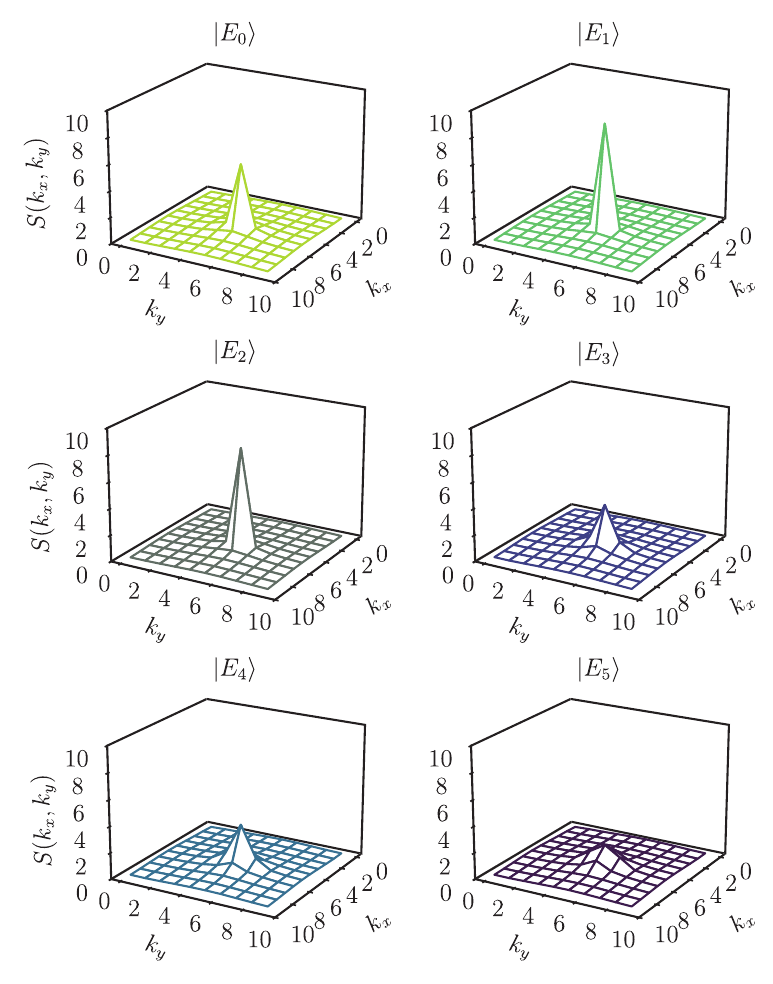}
    \caption{Spin structure factor $S(\boldsymbol{k})$ for the first 6 excited states of the Heisenberg model on the $10 \times 10$ lattice in the momentum $\boldsymbol{q}=\Gamma$ symmetry sector. 
    }
    \label{fig:10x10_ss}
\end{figure}

The quantum spin system we investigate is the paradigmatic antiferromagnetic Heisenberg model in two dimensions, whose Hamiltonian is:
\begin{equation}
\label{eq:hamiltonian}
    \hat{H} = \sum_{\langle i, j \rangle} \hat{\boldsymbol{S}}_{i} \cdot \hat{\boldsymbol{S}}_j,
\end{equation}
where $\hat{\boldsymbol{S}}_i = (\hat{S}_i^x, \hat{S}_i^y, \hat{S}_i^z)$ is the spin-$\frac{1}{2}$ operator at site $i$ and the sum runs over all the nearest-neighboring sites. 
We consider square lattice geometries with periodic boundary conditions along both spatial directions.

First, we benchmark on a $6 \times 6$ system, where exact diagonalization (ED) is still attainable. 
Besides the energies of the excited states, we also compute the spin structure factor:
\begin{equation}
    S(\boldsymbol{k}) = 
    \langle \hat{S}^\dagger_{\boldsymbol{k}} \hat{S}_{\boldsymbol{k}} \rangle \, \, \text{with} \, \, \hat{S}_{\boldsymbol{k}} = \frac{1}{\sqrt{N}} \sum_{\boldsymbol{l}} e^{-i \boldsymbol{k} \cdot \boldsymbol{l}} \hat{S}_{\boldsymbol{l}}^z, 
\end{equation}
where the vector $\boldsymbol{l}$ gives the position of the sites in the lattice. 
Due to the enforcement of both zero total $z$-magnetization and translational symmetry in the ansätze, the spin structure factor is equivalently the expectation of $\hat{S}^\dagger_{\boldsymbol{k}} \hat{P}_{\mathcal{V}^\perp} \hat{S}_{\boldsymbol{k}}$ for each of the basis wave functions that span $\mathcal{V}$.
We then obtain the  spin structure factor values from the diagonals of the OVM for $\hat{S}_{\boldsymbol{k}}$.

In \cref{fig:6x6}, we show the comparison between our results and the ED benchmark for the first 4 excited states in different momentum sectors. 
The results demonstrate remarkable accuracy in both the energy and the structure factor $S(\pi, \pi)$, across all excitations and momentum sectors. 
As shown in the inset, the relative energy errors remain consistently below or around $10^{-4}$ across the spectrum.
We observe that the discrepancies become more pronounced for higher excited states, in line with the findings of~\cite{pfau2024accurate}. 

After benchmarking, we investigate a larger $10 \times 10$ lattice. 
The corresponding 6 excited-state energies are shown in~\cref{fig:10x10}. 
The V-scores~\cite{wu2024variational} values reported in the inset validate the accuracy of the variational results, remaining of the order of $10^{-3}$ or below across all excited states and momentum sectors. 
We remark that for a given V-score the actual energy relative error is typically an order of magnitude smaller~\cite{wu2024variational}. 
For the ground state in the $\boldsymbol{q} = \Gamma$ sector, we also report the Quantum Monte Carlo (QMC) estimate for the energy~\cite{sandvik1997finite}, $E_{\text{QMC}}/N_s = -0.671549(4)$, which is in excellent agreement with our result $E/N_s = -0.671544(4)$. 
Finally, in~\cref{fig:10x10_ss}, we present three-dimensional plots of the spin structure factor in the $\boldsymbol{q} = \Gamma$ momentum sector. 
In all the cases, the peak occurs at $\boldsymbol{k} = (\pi, \pi)$, though with varying intensity across different excited states.
The peak value $S(\pi, \pi)=5.35(9)$ for the ground state agrees with the QMC result $S_{\text{QMC}}(\pi, \pi) = 5.3124(3)$~\cite{sandvik1997finite}.

For the simulations, we considered the following network hyperparameters: $G=3$ stages with $B=2, 2, 8$ number of blocks, $d = 4, 6, 12$ for the $6 \times 6$ and $d=8, 12, 20$ for the $10 \times 10$, $C=4$ for the $6 \times 6$ and $C = 6$ for the $10 \times 10$.
The filter sizes are set to $K=3$ and $K_{dw} = 5$, and the expansion ratio to $f=4$.
For the RBMs, we employed 16 hidden units for the $6 \times 6$ and 12 for the $10 \times 10$.
We typically use $2^{11}$ Monte Carlo samples and perform around $5000$ SR steps with a learning rate of $\eta = 0.15/N_s$.
The QGT inversion is stabilized with a diagonal shift of $10^{-3}$.
Finally, parameter updates are performed using the SPRING algorithm~\cite{goldshlager2024kaczmarz}.

\section{Conclusions}
\label{sec:conclusions}
We presented a rigorous geometric formulation of the variational principle for excited states introduced in~\cite{pfau2024accurate} in terms of Grassmannians in the Hilbert space. 
This enables the generalization of the Stochastic Reconfiguration method and the introduction of the multidimensional extensions of operator variances and subspace overlaps.
We apply the framework to investigate the properties of excited states in the 2D antiferromagnetic Heisenberg model, achieving accurate excitation energies and capturing key physical observables across a large portion of the spectrum. 
The approach overcomes limitations of prior methods — such as instability in dense spectra and reliance on symmetry constraints — without introducing additional penalty terms or fine-tuned hyperparameters.
Possible future directions include the application of the framework to lattice fermions and frustrated spin systems.
The generalization of the overlap to multiple wave functions opens interesting venues such as the simulation of real-time dynamics through subspace expansion methods~\cite{motta2024subspace,sinibaldi2024time} and schemes based on infidelity minimization~\cite{gutierrez2022real,jonsson2018neural,medvidovic2021classical,Donatella2023Infidelity,sinibaldi2023unbiasing,Gravina2024ptVMC}.
Potential future applications include integrating the method with quantum hardware through hybrid optimization. 

\section{Software}
All the simulations have been carried out using the software package NetKet~\cite{netket2,vicentini2022netket}. 
The code will be made public in a later revision of the manuscript.
We acknowledge the use of SpinED~\cite{spined} for the exact benchmarks. 

\section{Acknowledgments}
We warmly thank F. Vicentini for insightful discussions.  
A. S. is supported by SEFRI under Grant No.\ MB22.00051 (NEQS - Neural Quantum Simulation).
D. H. is supported by the EPFL Quantum Science and Engineering Post-Doctoral Fellowship.

\vspace{0.5 cm}
\emph{Note: Preliminary results from this work were presented at the 2024 APS March Meeting. 
During the preparation of the manuscript, we became aware of a related study by Ma et al.~\cite{ma2025solving}, in which the authors compute excited states using a similar framework in systems of trapped ions with long-range interactions. 
We also acknowledge a concurrent work by A. Khahn, L. Gravina, and F. Vicentini, where a similar formalism is discussed to describe subspaces and correct quantum dynamics.
This work is expected to appear simultaneously on the preprint server.}

\bibliography{bibliography}

\begin{thebibliography}{53}%
\makeatletter
\providecommand \@ifxundefined [1]{%
 \@ifx{#1\undefined}
}%
\providecommand \@ifnum [1]{%
 \ifnum #1\expandafter \@firstoftwo
 \else \expandafter \@secondoftwo
 \fi
}%
\providecommand \@ifx [1]{%
 \ifx #1\expandafter \@firstoftwo
 \else \expandafter \@secondoftwo
 \fi
}%
\providecommand \natexlab [1]{#1}%
\providecommand \enquote  [1]{``#1''}%
\providecommand \bibnamefont  [1]{#1}%
\providecommand \bibfnamefont [1]{#1}%
\providecommand \citenamefont [1]{#1}%
\providecommand \href@noop [0]{\@secondoftwo}%
\providecommand \href [0]{\begingroup \@sanitize@url \@href}%
\providecommand \@href[1]{\@@startlink{#1}\@@href}%
\providecommand \@@href[1]{\endgroup#1\@@endlink}%
\providecommand \@sanitize@url [0]{\catcode `\\12\catcode `\$12\catcode `\&12\catcode `\#12\catcode `\^12\catcode `\_12\catcode `\%12\relax}%
\providecommand \@@startlink[1]{}%
\providecommand \@@endlink[0]{}%
\providecommand \url  [0]{\begingroup\@sanitize@url \@url }%
\providecommand \@url [1]{\endgroup\@href {#1}{\urlprefix }}%
\providecommand \urlprefix  [0]{URL }%
\providecommand \Eprint [0]{\href }%
\providecommand \doibase [0]{https://doi.org/}%
\providecommand \selectlanguage [0]{\@gobble}%
\providecommand \bibinfo  [0]{\@secondoftwo}%
\providecommand \bibfield  [0]{\@secondoftwo}%
\providecommand \translation [1]{[#1]}%
\providecommand \BibitemOpen [0]{}%
\providecommand \bibitemStop [0]{}%
\providecommand \bibitemNoStop [0]{.\EOS\space}%
\providecommand \EOS [0]{\spacefactor3000\relax}%
\providecommand \BibitemShut  [1]{\csname bibitem#1\endcsname}%
\let\auto@bib@innerbib\@empty
\bibitem [{\citenamefont {Pfau}\ \emph {et~al.}(2024)\citenamefont {Pfau}, \citenamefont {Axelrod}, \citenamefont {Sutterud}, \citenamefont {von Glehn},\ and\ \citenamefont {Spencer}}]{pfau2024accurate}%
  \BibitemOpen
  \bibfield  {author} {\bibinfo {author} {\bibfnamefont {D.}~\bibnamefont {Pfau}}, \bibinfo {author} {\bibfnamefont {S.}~\bibnamefont {Axelrod}}, \bibinfo {author} {\bibfnamefont {H.}~\bibnamefont {Sutterud}}, \bibinfo {author} {\bibfnamefont {I.}~\bibnamefont {von Glehn}},\ and\ \bibinfo {author} {\bibfnamefont {J.~S.}\ \bibnamefont {Spencer}},\ }\href {https://doi.org/10.1126/science.adn0137} {\bibfield  {journal} {\bibinfo  {journal} {Science}\ }\textbf {\bibinfo {volume} {385}},\ \bibinfo {pages} {eadn0137} (\bibinfo {year} {2024})}\BibitemShut {NoStop}%
\bibitem [{\citenamefont {Stanton}\ and\ \citenamefont {Bartlett}(1993)}]{Stanton1993}%
  \BibitemOpen
  \bibfield  {author} {\bibinfo {author} {\bibfnamefont {J.~F.}\ \bibnamefont {Stanton}}\ and\ \bibinfo {author} {\bibfnamefont {R.~J.}\ \bibnamefont {Bartlett}},\ }\href {https://doi.org/10.1063/1.464746} {\bibfield  {journal} {\bibinfo  {journal} {The Journal of Chemical Physics}\ }\textbf {\bibinfo {volume} {98}},\ \bibinfo {pages} {7029} (\bibinfo {year} {1993})}\BibitemShut {NoStop}%
\bibitem [{\citenamefont {Troyer}\ and\ \citenamefont {Wiese}(2005)}]{Troyer2005}%
  \BibitemOpen
  \bibfield  {author} {\bibinfo {author} {\bibfnamefont {M.}~\bibnamefont {Troyer}}\ and\ \bibinfo {author} {\bibfnamefont {U.-J.}\ \bibnamefont {Wiese}},\ }\href {https://doi.org/10.1103/PhysRevLett.94.170201} {\bibfield  {journal} {\bibinfo  {journal} {Phys. Rev. Lett.}\ }\textbf {\bibinfo {volume} {94}},\ \bibinfo {pages} {170201} (\bibinfo {year} {2005})}\BibitemShut {NoStop}%
\bibitem [{\citenamefont {White}(1992)}]{white1992density}%
  \BibitemOpen
  \bibfield  {author} {\bibinfo {author} {\bibfnamefont {S.~R.}\ \bibnamefont {White}},\ }\href {https://doi.org/10.1103/PhysRevLett.69.2863} {\bibfield  {journal} {\bibinfo  {journal} {Phys. Rev. Lett.}\ }\textbf {\bibinfo {volume} {69}},\ \bibinfo {pages} {2863} (\bibinfo {year} {1992})}\BibitemShut {NoStop}%
\bibitem [{\citenamefont {Verstraete}\ and\ \citenamefont {Cirac}(2004)}]{Verstraete2004}%
  \BibitemOpen
  \bibfield  {author} {\bibinfo {author} {\bibfnamefont {F.}~\bibnamefont {Verstraete}}\ and\ \bibinfo {author} {\bibfnamefont {J.~I.}\ \bibnamefont {Cirac}},\ }\href@noop {} {\bibfield  {journal} {\bibinfo  {journal} {arXiv preprint}\ } (\bibinfo {year} {2004})},\ \Eprint {https://arxiv.org/abs/arXiv:cond-mat/0407066} {arXiv:cond-mat/0407066} \BibitemShut {NoStop}%
\bibitem [{\citenamefont {Vidal}(2007)}]{Vidal2007}%
  \BibitemOpen
  \bibfield  {author} {\bibinfo {author} {\bibfnamefont {G.}~\bibnamefont {Vidal}},\ }\href {https://doi.org/10.1103/PhysRevLett.99.220405} {\bibfield  {journal} {\bibinfo  {journal} {Phys. Rev. Lett.}\ }\textbf {\bibinfo {volume} {99}},\ \bibinfo {pages} {220405} (\bibinfo {year} {2007})}\BibitemShut {NoStop}%
\bibitem [{\citenamefont {Carleo}\ and\ \citenamefont {Troyer}(2017)}]{carleo2017solving}%
  \BibitemOpen
  \bibfield  {author} {\bibinfo {author} {\bibfnamefont {G.}~\bibnamefont {Carleo}}\ and\ \bibinfo {author} {\bibfnamefont {M.}~\bibnamefont {Troyer}},\ }\href {https://doi.org/10.1126/science.aag2302} {\bibfield  {journal} {\bibinfo  {journal} {Science}\ }\textbf {\bibinfo {volume} {355}},\ \bibinfo {pages} {602} (\bibinfo {year} {2017})}\BibitemShut {NoStop}%
\bibitem [{\citenamefont {Choo}\ \emph {et~al.}(2019)\citenamefont {Choo}, \citenamefont {Neupert},\ and\ \citenamefont {Carleo}}]{choo2019two}%
  \BibitemOpen
  \bibfield  {author} {\bibinfo {author} {\bibfnamefont {K.}~\bibnamefont {Choo}}, \bibinfo {author} {\bibfnamefont {T.}~\bibnamefont {Neupert}},\ and\ \bibinfo {author} {\bibfnamefont {G.}~\bibnamefont {Carleo}},\ }\href {https://doi.org/10.1103/PhysRevB.100.125124} {\bibfield  {journal} {\bibinfo  {journal} {Phys. Rev. B}\ }\textbf {\bibinfo {volume} {100}},\ \bibinfo {pages} {125124} (\bibinfo {year} {2019})}\BibitemShut {NoStop}%
\bibitem [{\citenamefont {Viteritti}\ \emph {et~al.}(2023)\citenamefont {Viteritti}, \citenamefont {Rende},\ and\ \citenamefont {Becca}}]{viteritti2023transformer_1d}%
  \BibitemOpen
  \bibfield  {author} {\bibinfo {author} {\bibfnamefont {L.~L.}\ \bibnamefont {Viteritti}}, \bibinfo {author} {\bibfnamefont {R.}~\bibnamefont {Rende}},\ and\ \bibinfo {author} {\bibfnamefont {F.}~\bibnamefont {Becca}},\ }\href {https://doi.org/10.1103/PhysRevLett.130.236401} {\bibfield  {journal} {\bibinfo  {journal} {Phys. Rev. Lett.}\ }\textbf {\bibinfo {volume} {130}},\ \bibinfo {pages} {236401} (\bibinfo {year} {2023})}\BibitemShut {NoStop}%
\bibitem [{\citenamefont {Viteritti}\ \emph {et~al.}(2025)\citenamefont {Viteritti}, \citenamefont {Rende}, \citenamefont {Parola}, \citenamefont {Goldt},\ and\ \citenamefont {Becca}}]{viteritti2023transformer_2d}%
  \BibitemOpen
  \bibfield  {author} {\bibinfo {author} {\bibfnamefont {L.~L.}\ \bibnamefont {Viteritti}}, \bibinfo {author} {\bibfnamefont {R.}~\bibnamefont {Rende}}, \bibinfo {author} {\bibfnamefont {A.}~\bibnamefont {Parola}}, \bibinfo {author} {\bibfnamefont {S.}~\bibnamefont {Goldt}},\ and\ \bibinfo {author} {\bibfnamefont {F.}~\bibnamefont {Becca}},\ }\href {https://doi.org/10.1103/PhysRevB.111.134411} {\bibfield  {journal} {\bibinfo  {journal} {Phys. Rev. B}\ }\textbf {\bibinfo {volume} {111}},\ \bibinfo {pages} {134411} (\bibinfo {year} {2025})}\BibitemShut {NoStop}%
\bibitem [{\citenamefont {Chen}\ and\ \citenamefont {Heyl}(2024)}]{chen2024empowering}%
  \BibitemOpen
  \bibfield  {author} {\bibinfo {author} {\bibfnamefont {A.}~\bibnamefont {Chen}}\ and\ \bibinfo {author} {\bibfnamefont {M.}~\bibnamefont {Heyl}},\ }\href {https://doi.org/10.1038/s41567-024-02566-1} {\bibfield  {journal} {\bibinfo  {journal} {Nature Physics}\ }\textbf {\bibinfo {volume} {20}},\ \bibinfo {pages} {1476} (\bibinfo {year} {2024})}\BibitemShut {NoStop}%
\bibitem [{\citenamefont {Denis}\ and\ \citenamefont {Carleo}(2025)}]{Denis2025accurateneural}%
  \BibitemOpen
  \bibfield  {author} {\bibinfo {author} {\bibfnamefont {Z.}~\bibnamefont {Denis}}\ and\ \bibinfo {author} {\bibfnamefont {G.}~\bibnamefont {Carleo}},\ }\href {https://doi.org/10.22331/q-2025-06-17-1772} {\bibfield  {journal} {\bibinfo  {journal} {{Quantum}}\ }\textbf {\bibinfo {volume} {9}},\ \bibinfo {pages} {1772} (\bibinfo {year} {2025})}\BibitemShut {NoStop}%
\bibitem [{\citenamefont {Choo}\ \emph {et~al.}(2020)\citenamefont {Choo}, \citenamefont {Mezzacapo},\ and\ \citenamefont {Carleo}}]{Choo2020Fermions}%
  \BibitemOpen
  \bibfield  {author} {\bibinfo {author} {\bibfnamefont {K.}~\bibnamefont {Choo}}, \bibinfo {author} {\bibfnamefont {A.}~\bibnamefont {Mezzacapo}},\ and\ \bibinfo {author} {\bibfnamefont {G.}~\bibnamefont {Carleo}},\ }\bibfield  {journal} {\bibinfo  {journal} {Nature Communications}\ }\textbf {\bibinfo {volume} {11}},\ \href {https://doi.org/10.1038/s41467-020-15724-9} {10.1038/s41467-020-15724-9} (\bibinfo {year} {2020})\BibitemShut {NoStop}%
\bibitem [{\citenamefont {Pfau}\ \emph {et~al.}(2020)\citenamefont {Pfau}, \citenamefont {Spencer}, \citenamefont {Matthews},\ and\ \citenamefont {Foulkes}}]{Pfau2020PRR}%
  \BibitemOpen
  \bibfield  {author} {\bibinfo {author} {\bibfnamefont {D.}~\bibnamefont {Pfau}}, \bibinfo {author} {\bibfnamefont {J.~S.}\ \bibnamefont {Spencer}}, \bibinfo {author} {\bibfnamefont {A.~G. D.~G.}\ \bibnamefont {Matthews}},\ and\ \bibinfo {author} {\bibfnamefont {W.~M.~C.}\ \bibnamefont {Foulkes}},\ }\href {https://doi.org/10.1103/PhysRevResearch.2.033429} {\bibfield  {journal} {\bibinfo  {journal} {Phys. Rev. Research}\ }\textbf {\bibinfo {volume} {2}},\ \bibinfo {pages} {033429} (\bibinfo {year} {2020})}\BibitemShut {NoStop}%
\bibitem [{\citenamefont {Nomura}\ \emph {et~al.}(2017)\citenamefont {Nomura}, \citenamefont {Darmawan}, \citenamefont {Yamaji},\ and\ \citenamefont {Imada}}]{NomuraPRBFermionic}%
  \BibitemOpen
  \bibfield  {author} {\bibinfo {author} {\bibfnamefont {Y.}~\bibnamefont {Nomura}}, \bibinfo {author} {\bibfnamefont {A.~S.}\ \bibnamefont {Darmawan}}, \bibinfo {author} {\bibfnamefont {Y.}~\bibnamefont {Yamaji}},\ and\ \bibinfo {author} {\bibfnamefont {M.}~\bibnamefont {Imada}},\ }\href {https://doi.org/10.1103/PhysRevB.96.205152} {\bibfield  {journal} {\bibinfo  {journal} {Phys. Rev. B}\ }\textbf {\bibinfo {volume} {96}},\ \bibinfo {pages} {205152} (\bibinfo {year} {2017})}\BibitemShut {NoStop}%
\bibitem [{\citenamefont {Stokes}\ \emph {et~al.}(2020{\natexlab{a}})\citenamefont {Stokes}, \citenamefont {Moreno}, \citenamefont {Pnevmatikakis},\ and\ \citenamefont {Carleo}}]{Stokes2020PRB}%
  \BibitemOpen
  \bibfield  {author} {\bibinfo {author} {\bibfnamefont {J.}~\bibnamefont {Stokes}}, \bibinfo {author} {\bibfnamefont {J.~R.}\ \bibnamefont {Moreno}}, \bibinfo {author} {\bibfnamefont {E.~A.}\ \bibnamefont {Pnevmatikakis}},\ and\ \bibinfo {author} {\bibfnamefont {G.}~\bibnamefont {Carleo}},\ }\href {https://doi.org/10.1103/PhysRevB.102.205122} {\bibfield  {journal} {\bibinfo  {journal} {Phys. Rev. B}\ }\textbf {\bibinfo {volume} {102}},\ \bibinfo {pages} {205122} (\bibinfo {year} {2020}{\natexlab{a}})}\BibitemShut {NoStop}%
\bibitem [{\citenamefont {Nys}\ and\ \citenamefont {Carleo}(2022)}]{Nys22Fermions}%
  \BibitemOpen
  \bibfield  {author} {\bibinfo {author} {\bibfnamefont {J.}~\bibnamefont {Nys}}\ and\ \bibinfo {author} {\bibfnamefont {G.}~\bibnamefont {Carleo}},\ }\href {https://doi.org/10.22331/q-2022-10-13-833} {\bibfield  {journal} {\bibinfo  {journal} {{Quantum}}\ }\textbf {\bibinfo {volume} {6}},\ \bibinfo {pages} {833} (\bibinfo {year} {2022})}\BibitemShut {NoStop}%
\bibitem [{\citenamefont {Schmitt}\ and\ \citenamefont {Heyl}(2020)}]{schmitt2020quantum}%
  \BibitemOpen
  \bibfield  {author} {\bibinfo {author} {\bibfnamefont {M.}~\bibnamefont {Schmitt}}\ and\ \bibinfo {author} {\bibfnamefont {M.}~\bibnamefont {Heyl}},\ }\href {https://doi.org/10.1103/PhysRevLett.125.100503} {\bibfield  {journal} {\bibinfo  {journal} {Phys. Rev. Lett.}\ }\textbf {\bibinfo {volume} {125}},\ \bibinfo {pages} {100503} (\bibinfo {year} {2020})}\BibitemShut {NoStop}%
\bibitem [{\citenamefont {Sinibaldi}\ \emph {et~al.}(2023)\citenamefont {Sinibaldi}, \citenamefont {Giuliani}, \citenamefont {Carleo},\ and\ \citenamefont {Vicentini}}]{sinibaldi2023unbiasing}%
  \BibitemOpen
  \bibfield  {author} {\bibinfo {author} {\bibfnamefont {A.}~\bibnamefont {Sinibaldi}}, \bibinfo {author} {\bibfnamefont {C.}~\bibnamefont {Giuliani}}, \bibinfo {author} {\bibfnamefont {G.}~\bibnamefont {Carleo}},\ and\ \bibinfo {author} {\bibfnamefont {F.}~\bibnamefont {Vicentini}},\ }\href {https://doi.org/10.22331/q-2023-10-10-1131} {\bibfield  {journal} {\bibinfo  {journal} {{Quantum}}\ }\textbf {\bibinfo {volume} {7}},\ \bibinfo {pages} {1131} (\bibinfo {year} {2023})}\BibitemShut {NoStop}%
\bibitem [{\citenamefont {Nys}\ \emph {et~al.}(2024)\citenamefont {Nys}, \citenamefont {Pescia}, \citenamefont {Sinibaldi},\ and\ \citenamefont {Carleo}}]{nys2024ab}%
  \BibitemOpen
  \bibfield  {author} {\bibinfo {author} {\bibfnamefont {J.}~\bibnamefont {Nys}}, \bibinfo {author} {\bibfnamefont {G.}~\bibnamefont {Pescia}}, \bibinfo {author} {\bibfnamefont {A.}~\bibnamefont {Sinibaldi}},\ and\ \bibinfo {author} {\bibfnamefont {G.}~\bibnamefont {Carleo}},\ }\href {https://doi.org/10.1038/s41467-024-53672-w} {\bibfield  {journal} {\bibinfo  {journal} {Nature Communications}\ }\textbf {\bibinfo {volume} {15}},\ \bibinfo {pages} {9404} (\bibinfo {year} {2024})}\BibitemShut {NoStop}%
\bibitem [{\citenamefont {Sinibaldi}\ \emph {et~al.}(2024)\citenamefont {Sinibaldi}, \citenamefont {Hendry}, \citenamefont {Vicentini},\ and\ \citenamefont {Carleo}}]{sinibaldi2024time}%
  \BibitemOpen
  \bibfield  {author} {\bibinfo {author} {\bibfnamefont {A.}~\bibnamefont {Sinibaldi}}, \bibinfo {author} {\bibfnamefont {D.}~\bibnamefont {Hendry}}, \bibinfo {author} {\bibfnamefont {F.}~\bibnamefont {Vicentini}},\ and\ \bibinfo {author} {\bibfnamefont {G.}~\bibnamefont {Carleo}},\ }\href@noop {} {\bibfield  {journal} {\bibinfo  {journal} {arXiv preprint}\ } (\bibinfo {year} {2024})},\ \Eprint {https://arxiv.org/abs/arXiv:2412.11778} {arXiv:2412.11778} \BibitemShut {NoStop}%
\bibitem [{\citenamefont {Choo}\ \emph {et~al.}(2018)\citenamefont {Choo}, \citenamefont {Carleo}, \citenamefont {Regnault},\ and\ \citenamefont {Neupert}}]{choo_excited}%
  \BibitemOpen
  \bibfield  {author} {\bibinfo {author} {\bibfnamefont {K.}~\bibnamefont {Choo}}, \bibinfo {author} {\bibfnamefont {G.}~\bibnamefont {Carleo}}, \bibinfo {author} {\bibfnamefont {N.}~\bibnamefont {Regnault}},\ and\ \bibinfo {author} {\bibfnamefont {T.}~\bibnamefont {Neupert}},\ }\href {https://doi.org/10.1103/PhysRevLett.121.167204} {\bibfield  {journal} {\bibinfo  {journal} {Phys. Rev. Lett.}\ }\textbf {\bibinfo {volume} {121}},\ \bibinfo {pages} {167204} (\bibinfo {year} {2018})}\BibitemShut {NoStop}%
\bibitem [{\citenamefont {Entwistle}\ \emph {et~al.}(2023)\citenamefont {Entwistle}, \citenamefont {Sch{\"a}tzle}, \citenamefont {Erdman}, \citenamefont {Hermann},\ and\ \citenamefont {No{\'e}}}]{entwistle2023electronic}%
  \BibitemOpen
  \bibfield  {author} {\bibinfo {author} {\bibfnamefont {M.~T.}\ \bibnamefont {Entwistle}}, \bibinfo {author} {\bibfnamefont {Z.}~\bibnamefont {Sch{\"a}tzle}}, \bibinfo {author} {\bibfnamefont {P.~A.}\ \bibnamefont {Erdman}}, \bibinfo {author} {\bibfnamefont {J.}~\bibnamefont {Hermann}},\ and\ \bibinfo {author} {\bibfnamefont {F.}~\bibnamefont {No{\'e}}},\ }\href {https://doi.org/https://doi.org/10.1038/s41467-022-35534-5} {\bibfield  {journal} {\bibinfo  {journal} {Nature Communications}\ }\textbf {\bibinfo {volume} {14}},\ \bibinfo {pages} {274} (\bibinfo {year} {2023})}\BibitemShut {NoStop}%
\bibitem [{\citenamefont {Pathak}\ \emph {et~al.}(2021)\citenamefont {Pathak}, \citenamefont {Busemeyer}, \citenamefont {Rodrigues},\ and\ \citenamefont {Wagner}}]{pathak2021excited}%
  \BibitemOpen
  \bibfield  {author} {\bibinfo {author} {\bibfnamefont {S.}~\bibnamefont {Pathak}}, \bibinfo {author} {\bibfnamefont {B.}~\bibnamefont {Busemeyer}}, \bibinfo {author} {\bibfnamefont {J.~N.}\ \bibnamefont {Rodrigues}},\ and\ \bibinfo {author} {\bibfnamefont {L.~K.}\ \bibnamefont {Wagner}},\ }\href {https://doi.org/https://doi.org/10.1063/5.0032172} {\bibfield  {journal} {\bibinfo  {journal} {The Journal of Chemical Physics}\ }\textbf {\bibinfo {volume} {154}} (\bibinfo {year} {2021})}\BibitemShut {NoStop}%
\bibitem [{\citenamefont {McMillan}(1965)}]{mcmillan1965ground}%
  \BibitemOpen
  \bibfield  {author} {\bibinfo {author} {\bibfnamefont {W.~L.}\ \bibnamefont {McMillan}},\ }\href {https://doi.org/10.1103/PhysRev.138.A442} {\bibfield  {journal} {\bibinfo  {journal} {Phys. Rev.}\ }\textbf {\bibinfo {volume} {138}},\ \bibinfo {pages} {A442} (\bibinfo {year} {1965})}\BibitemShut {NoStop}%
\bibitem [{\citenamefont {Sorella}(1998)}]{sorella1998green}%
  \BibitemOpen
  \bibfield  {author} {\bibinfo {author} {\bibfnamefont {S.}~\bibnamefont {Sorella}},\ }\href {https://doi.org/10.1103/PhysRevLett.80.4558} {\bibfield  {journal} {\bibinfo  {journal} {Phys. Rev. Lett.}\ }\textbf {\bibinfo {volume} {80}},\ \bibinfo {pages} {4558} (\bibinfo {year} {1998})}\BibitemShut {NoStop}%
\bibitem [{\citenamefont {Absil}\ \emph {et~al.}(2004)\citenamefont {Absil}, \citenamefont {Mahony},\ and\ \citenamefont {Sepulchre}}]{absil2004riemannian}%
  \BibitemOpen
  \bibfield  {author} {\bibinfo {author} {\bibfnamefont {P.-A.}\ \bibnamefont {Absil}}, \bibinfo {author} {\bibfnamefont {R.}~\bibnamefont {Mahony}},\ and\ \bibinfo {author} {\bibfnamefont {R.}~\bibnamefont {Sepulchre}},\ }\href {https://doi.org/10.1023/B:ACAP.0000013855.14971.91} {\bibfield  {journal} {\bibinfo  {journal} {Acta Applicandae Mathematica}\ }\textbf {\bibinfo {volume} {80}},\ \bibinfo {pages} {199} (\bibinfo {year} {2004})}\BibitemShut {NoStop}%
\bibitem [{\citenamefont {Hackl}\ \emph {et~al.}(2020)\citenamefont {Hackl}, \citenamefont {Guaita}, \citenamefont {Shi}, \citenamefont {Haegeman}, \citenamefont {Demler},\ and\ \citenamefont {Cirac}}]{hackl2020geometry}%
  \BibitemOpen
  \bibfield  {author} {\bibinfo {author} {\bibfnamefont {L.}~\bibnamefont {Hackl}}, \bibinfo {author} {\bibfnamefont {T.}~\bibnamefont {Guaita}}, \bibinfo {author} {\bibfnamefont {T.}~\bibnamefont {Shi}}, \bibinfo {author} {\bibfnamefont {J.}~\bibnamefont {Haegeman}}, \bibinfo {author} {\bibfnamefont {E.}~\bibnamefont {Demler}},\ and\ \bibinfo {author} {\bibfnamefont {J.~I.}\ \bibnamefont {Cirac}},\ }\href {https://doi.org/10.21468/SciPostPhys.9.4.048} {\bibfield  {journal} {\bibinfo  {journal} {SciPost Phys.}\ }\textbf {\bibinfo {volume} {9}},\ \bibinfo {pages} {048} (\bibinfo {year} {2020})}\BibitemShut {NoStop}%
\bibitem [{\citenamefont {Luo}\ and\ \citenamefont {Clark}(2019)}]{diluo2019}%
  \BibitemOpen
  \bibfield  {author} {\bibinfo {author} {\bibfnamefont {D.}~\bibnamefont {Luo}}\ and\ \bibinfo {author} {\bibfnamefont {B.~K.}\ \bibnamefont {Clark}},\ }\href {https://doi.org/10.1103/PhysRevLett.122.226401} {\bibfield  {journal} {\bibinfo  {journal} {Phys. Rev. Lett.}\ }\textbf {\bibinfo {volume} {122}},\ \bibinfo {pages} {226401} (\bibinfo {year} {2019})}\BibitemShut {NoStop}%
\bibitem [{\citenamefont {Stokes}\ \emph {et~al.}(2020{\natexlab{b}})\citenamefont {Stokes}, \citenamefont {Izaac}, \citenamefont {Killoran},\ and\ \citenamefont {Carleo}}]{stokes2020quantum}%
  \BibitemOpen
  \bibfield  {author} {\bibinfo {author} {\bibfnamefont {J.}~\bibnamefont {Stokes}}, \bibinfo {author} {\bibfnamefont {J.}~\bibnamefont {Izaac}}, \bibinfo {author} {\bibfnamefont {N.}~\bibnamefont {Killoran}},\ and\ \bibinfo {author} {\bibfnamefont {G.}~\bibnamefont {Carleo}},\ }\href {https://doi.org/10.22331/q-2020-05-25-269} {\bibfield  {journal} {\bibinfo  {journal} {{Quantum}}\ }\textbf {\bibinfo {volume} {4}},\ \bibinfo {pages} {269} (\bibinfo {year} {2020}{\natexlab{b}})}\BibitemShut {NoStop}%
\bibitem [{\citenamefont {Park}\ and\ \citenamefont {Kastoryano}(2020)}]{park2020geometry}%
  \BibitemOpen
  \bibfield  {author} {\bibinfo {author} {\bibfnamefont {C.-Y.}\ \bibnamefont {Park}}\ and\ \bibinfo {author} {\bibfnamefont {M.~J.}\ \bibnamefont {Kastoryano}},\ }\href {https://doi.org/10.1103/PhysRevResearch.2.023232} {\bibfield  {journal} {\bibinfo  {journal} {Phys. Rev. Res.}\ }\textbf {\bibinfo {volume} {2}},\ \bibinfo {pages} {023232} (\bibinfo {year} {2020})}\BibitemShut {NoStop}%
\bibitem [{\citenamefont {Liu}\ \emph {et~al.}(2022)\citenamefont {Liu}, \citenamefont {Mao}, \citenamefont {Wu}, \citenamefont {Feichtenhofer}, \citenamefont {Darrell},\ and\ \citenamefont {Xie}}]{liu2022convnet}%
  \BibitemOpen
  \bibfield  {author} {\bibinfo {author} {\bibfnamefont {Z.}~\bibnamefont {Liu}}, \bibinfo {author} {\bibfnamefont {H.}~\bibnamefont {Mao}}, \bibinfo {author} {\bibfnamefont {C.-Y.}\ \bibnamefont {Wu}}, \bibinfo {author} {\bibfnamefont {C.}~\bibnamefont {Feichtenhofer}}, \bibinfo {author} {\bibfnamefont {T.}~\bibnamefont {Darrell}},\ and\ \bibinfo {author} {\bibfnamefont {S.}~\bibnamefont {Xie}},\ }in\ \href {https://doi.org/10.1109/CVPR52688.2022.01167} {\emph {\bibinfo {booktitle} {Proceedings of the IEEE/CVF conference on computer vision and pattern recognition}}}\ (\bibinfo {year} {2022})\ pp.\ \bibinfo {pages} {11976--11986}\BibitemShut {NoStop}%
\bibitem [{\citenamefont {Woo}\ \emph {et~al.}(2023)\citenamefont {Woo}, \citenamefont {Debnath}, \citenamefont {Hu}, \citenamefont {Chen}, \citenamefont {Liu}, \citenamefont {Kweon},\ and\ \citenamefont {Xie}}]{woo2023convnext}%
  \BibitemOpen
  \bibfield  {author} {\bibinfo {author} {\bibfnamefont {S.}~\bibnamefont {Woo}}, \bibinfo {author} {\bibfnamefont {S.}~\bibnamefont {Debnath}}, \bibinfo {author} {\bibfnamefont {R.}~\bibnamefont {Hu}}, \bibinfo {author} {\bibfnamefont {X.}~\bibnamefont {Chen}}, \bibinfo {author} {\bibfnamefont {Z.}~\bibnamefont {Liu}}, \bibinfo {author} {\bibfnamefont {I.~S.}\ \bibnamefont {Kweon}},\ and\ \bibinfo {author} {\bibfnamefont {S.}~\bibnamefont {Xie}},\ }in\ \href {https://doi.org/10.1109/CVPR52729.2023.01548} {\emph {\bibinfo {booktitle} {Proceedings of the IEEE/CVF conference on computer vision and pattern recognition}}}\ (\bibinfo {year} {2023})\ pp.\ \bibinfo {pages} {16133--16142}\BibitemShut {NoStop}%
\bibitem [{\citenamefont {Sinibaldi}\ \emph {et~al.}(2025)\citenamefont {Sinibaldi}, \citenamefont {Mello}, \citenamefont {Collura},\ and\ \citenamefont {Carleo}}]{sinibaldi2025non}%
  \BibitemOpen
  \bibfield  {author} {\bibinfo {author} {\bibfnamefont {A.}~\bibnamefont {Sinibaldi}}, \bibinfo {author} {\bibfnamefont {A.~F.}\ \bibnamefont {Mello}}, \bibinfo {author} {\bibfnamefont {M.}~\bibnamefont {Collura}},\ and\ \bibinfo {author} {\bibfnamefont {G.}~\bibnamefont {Carleo}},\ }\href@noop {} {\bibfield  {journal} {\bibinfo  {journal} {arXiv preprint}\ } (\bibinfo {year} {2025})},\ \Eprint {https://arxiv.org/abs/arXiv:2502.09725} {arXiv:2502.09725} \BibitemShut {NoStop}%
\bibitem [{\citenamefont {Nutakki}\ \emph {et~al.}(2025)\citenamefont {Nutakki}, \citenamefont {Shokry},\ and\ \citenamefont {Vicentini}}]{nutakki2025design}%
  \BibitemOpen
  \bibfield  {author} {\bibinfo {author} {\bibfnamefont {R.~P.}\ \bibnamefont {Nutakki}}, \bibinfo {author} {\bibfnamefont {A.}~\bibnamefont {Shokry}},\ and\ \bibinfo {author} {\bibfnamefont {F.}~\bibnamefont {Vicentini}},\ }\href@noop {} {\bibfield  {journal} {\bibinfo  {journal} {arXiv preprint}\ } (\bibinfo {year} {2025})},\ \Eprint {https://arxiv.org/abs/arXiv:2505.03466} {arXiv:2505.03466} \BibitemShut {NoStop}%
\bibitem [{\citenamefont {Ba}\ \emph {et~al.}(2016)\citenamefont {Ba}, \citenamefont {Kiros},\ and\ \citenamefont {Hinton}}]{ba2016layer}%
  \BibitemOpen
  \bibfield  {author} {\bibinfo {author} {\bibfnamefont {J.~L.}\ \bibnamefont {Ba}}, \bibinfo {author} {\bibfnamefont {J.~R.}\ \bibnamefont {Kiros}},\ and\ \bibinfo {author} {\bibfnamefont {G.~E.}\ \bibnamefont {Hinton}},\ }\href@noop {} {\bibfield  {journal} {\bibinfo  {journal} {arXiv preprint}\ } (\bibinfo {year} {2016})},\ \Eprint {https://arxiv.org/abs/arXiv:1607.06450} {arXiv:1607.06450} \BibitemShut {NoStop}%
\bibitem [{\citenamefont {Bengio}\ \emph {et~al.}(2013)\citenamefont {Bengio}, \citenamefont {Courville},\ and\ \citenamefont {Vincent}}]{bengio2013representation}%
  \BibitemOpen
  \bibfield  {author} {\bibinfo {author} {\bibfnamefont {Y.}~\bibnamefont {Bengio}}, \bibinfo {author} {\bibfnamefont {A.}~\bibnamefont {Courville}},\ and\ \bibinfo {author} {\bibfnamefont {P.}~\bibnamefont {Vincent}},\ }\href {https://doi.org/10.1016/j.rcim.2025.103011} {\bibfield  {journal} {\bibinfo  {journal} {IEEE transactions on pattern analysis and machine intelligence}\ }\textbf {\bibinfo {volume} {35}},\ \bibinfo {pages} {1798} (\bibinfo {year} {2013})}\BibitemShut {NoStop}%
\bibitem [{\citenamefont {Denis}\ and\ \citenamefont {Carleo}(2024)}]{denis2024accurate}%
  \BibitemOpen
  \bibfield  {author} {\bibinfo {author} {\bibfnamefont {Z.}~\bibnamefont {Denis}}\ and\ \bibinfo {author} {\bibfnamefont {G.}~\bibnamefont {Carleo}},\ }\href@noop {} {\bibfield  {journal} {\bibinfo  {journal} {arXiv preprint}\ } (\bibinfo {year} {2024})},\ \Eprint {https://arxiv.org/abs/arXiv:2404.07869} {arXiv:2404.07869} \BibitemShut {NoStop}%
\bibitem [{\citenamefont {Rende}\ \emph {et~al.}(2024)\citenamefont {Rende}, \citenamefont {Viteritti}, \citenamefont {Bardone}, \citenamefont {Becca},\ and\ \citenamefont {Goldt}}]{rende2024simple}%
  \BibitemOpen
  \bibfield  {author} {\bibinfo {author} {\bibfnamefont {R.}~\bibnamefont {Rende}}, \bibinfo {author} {\bibfnamefont {L.~L.}\ \bibnamefont {Viteritti}}, \bibinfo {author} {\bibfnamefont {L.}~\bibnamefont {Bardone}}, \bibinfo {author} {\bibfnamefont {F.}~\bibnamefont {Becca}},\ and\ \bibinfo {author} {\bibfnamefont {S.}~\bibnamefont {Goldt}},\ }\href {https://doi.org/10.1038/s42005-024-01732-4} {\bibfield  {journal} {\bibinfo  {journal} {Communications Physics}\ }\textbf {\bibinfo {volume} {7}},\ \bibinfo {pages} {260} (\bibinfo {year} {2024})}\BibitemShut {NoStop}%
\bibitem [{\citenamefont {Marshall}(1955)}]{marshall1955antiferromagnetism}%
  \BibitemOpen
  \bibfield  {author} {\bibinfo {author} {\bibfnamefont {W.}~\bibnamefont {Marshall}},\ }\href {https://doi.org/10.1098/rspa.1955.0200} {\bibfield  {journal} {\bibinfo  {journal} {Proceedings of the Royal Society of London. Series A. Mathematical and Physical Sciences}\ }\textbf {\bibinfo {volume} {232}},\ \bibinfo {pages} {48} (\bibinfo {year} {1955})}\BibitemShut {NoStop}%
\bibitem [{\citenamefont {Sandvik}(1997)}]{sandvik1997finite}%
  \BibitemOpen
  \bibfield  {author} {\bibinfo {author} {\bibfnamefont {A.~W.}\ \bibnamefont {Sandvik}},\ }\href {https://doi.org/10.1103/PhysRevB.56.11678} {\bibfield  {journal} {\bibinfo  {journal} {Phys. Rev. B}\ }\textbf {\bibinfo {volume} {56}},\ \bibinfo {pages} {11678} (\bibinfo {year} {1997})}\BibitemShut {NoStop}%
\bibitem [{\citenamefont {Wu}\ \emph {et~al.}(2024)\citenamefont {Wu}, \citenamefont {Rossi}, \citenamefont {Vicentini}, \citenamefont {Astrakhantsev}, \citenamefont {Becca}, \citenamefont {Cao}, \citenamefont {Carrasquilla}, \citenamefont {Ferrari}, \citenamefont {Georges}, \citenamefont {Hibat-Allah} \emph {et~al.}}]{wu2024variational}%
  \BibitemOpen
  \bibfield  {author} {\bibinfo {author} {\bibfnamefont {D.}~\bibnamefont {Wu}}, \bibinfo {author} {\bibfnamefont {R.}~\bibnamefont {Rossi}}, \bibinfo {author} {\bibfnamefont {F.}~\bibnamefont {Vicentini}}, \bibinfo {author} {\bibfnamefont {N.}~\bibnamefont {Astrakhantsev}}, \bibinfo {author} {\bibfnamefont {F.}~\bibnamefont {Becca}}, \bibinfo {author} {\bibfnamefont {X.}~\bibnamefont {Cao}}, \bibinfo {author} {\bibfnamefont {J.}~\bibnamefont {Carrasquilla}}, \bibinfo {author} {\bibfnamefont {F.}~\bibnamefont {Ferrari}}, \bibinfo {author} {\bibfnamefont {A.}~\bibnamefont {Georges}}, \bibinfo {author} {\bibfnamefont {M.}~\bibnamefont {Hibat-Allah}}, \emph {et~al.},\ }\href {https://doi.org/10.1126/science.adg9774} {\bibfield  {journal} {\bibinfo  {journal} {Science}\ }\textbf {\bibinfo {volume} {386}},\ \bibinfo {pages} {296} (\bibinfo {year} {2024})}\BibitemShut {NoStop}%
\bibitem [{\citenamefont {Goldshlager}\ \emph {et~al.}(2024)\citenamefont {Goldshlager}, \citenamefont {Abrahamsen},\ and\ \citenamefont {Lin}}]{goldshlager2024kaczmarz}%
  \BibitemOpen
  \bibfield  {author} {\bibinfo {author} {\bibfnamefont {G.}~\bibnamefont {Goldshlager}}, \bibinfo {author} {\bibfnamefont {N.}~\bibnamefont {Abrahamsen}},\ and\ \bibinfo {author} {\bibfnamefont {L.}~\bibnamefont {Lin}},\ }\href {https://doi.org/10.1016/j.jcp.2024.113351} {\bibfield  {journal} {\bibinfo  {journal} {Journal of Computational Physics}\ }\textbf {\bibinfo {volume} {516}},\ \bibinfo {pages} {113351} (\bibinfo {year} {2024})}\BibitemShut {NoStop}%
\bibitem [{\citenamefont {Motta}\ \emph {et~al.}(2024)\citenamefont {Motta}, \citenamefont {Kirby}, \citenamefont {Liepuoniute}, \citenamefont {Sung}, \citenamefont {Cohn}, \citenamefont {Mezzacapo}, \citenamefont {Klymko}, \citenamefont {Nguyen}, \citenamefont {Yoshioka},\ and\ \citenamefont {Rice}}]{motta2024subspace}%
  \BibitemOpen
  \bibfield  {author} {\bibinfo {author} {\bibfnamefont {M.}~\bibnamefont {Motta}}, \bibinfo {author} {\bibfnamefont {W.}~\bibnamefont {Kirby}}, \bibinfo {author} {\bibfnamefont {I.}~\bibnamefont {Liepuoniute}}, \bibinfo {author} {\bibfnamefont {K.~J.}\ \bibnamefont {Sung}}, \bibinfo {author} {\bibfnamefont {J.}~\bibnamefont {Cohn}}, \bibinfo {author} {\bibfnamefont {A.}~\bibnamefont {Mezzacapo}}, \bibinfo {author} {\bibfnamefont {K.}~\bibnamefont {Klymko}}, \bibinfo {author} {\bibfnamefont {N.}~\bibnamefont {Nguyen}}, \bibinfo {author} {\bibfnamefont {N.}~\bibnamefont {Yoshioka}},\ and\ \bibinfo {author} {\bibfnamefont {J.~E.}\ \bibnamefont {Rice}},\ }\href {https://doi.org/10.1088/2516-1075/ad3592} {\bibfield  {journal} {\bibinfo  {journal} {Electronic Structure}\ }\textbf {\bibinfo {volume} {6}},\ \bibinfo {pages} {013001} (\bibinfo {year} {2024})}\BibitemShut {NoStop}%
\bibitem [{\citenamefont {Guti{\'{e}}rrez}\ and\ \citenamefont {Mendl}(2022)}]{gutierrez2022real}%
  \BibitemOpen
  \bibfield  {author} {\bibinfo {author} {\bibfnamefont {I.~L.}\ \bibnamefont {Guti{\'{e}}rrez}}\ and\ \bibinfo {author} {\bibfnamefont {C.~B.}\ \bibnamefont {Mendl}},\ }\href {https://doi.org/10.22331/q-2022-01-20-627} {\bibfield  {journal} {\bibinfo  {journal} {{Quantum}}\ }\textbf {\bibinfo {volume} {6}},\ \bibinfo {pages} {627} (\bibinfo {year} {2022})}\BibitemShut {NoStop}%
\bibitem [{\citenamefont {J{\'o}nsson}\ \emph {et~al.}(2018)\citenamefont {J{\'o}nsson}, \citenamefont {Bauer},\ and\ \citenamefont {Carleo}}]{jonsson2018neural}%
  \BibitemOpen
  \bibfield  {author} {\bibinfo {author} {\bibfnamefont {B.}~\bibnamefont {J{\'o}nsson}}, \bibinfo {author} {\bibfnamefont {B.}~\bibnamefont {Bauer}},\ and\ \bibinfo {author} {\bibfnamefont {G.}~\bibnamefont {Carleo}},\ }\href@noop {} {\bibfield  {journal} {\bibinfo  {journal} {arXiv preprint}\ } (\bibinfo {year} {2018})},\ \Eprint {https://arxiv.org/abs/arXiv:1808.05232} {arXiv:1808.05232} \BibitemShut {NoStop}%
\bibitem [{\citenamefont {Medvidovi{\'c}}\ and\ \citenamefont {Carleo}(2021)}]{medvidovic2021classical}%
  \BibitemOpen
  \bibfield  {author} {\bibinfo {author} {\bibfnamefont {M.}~\bibnamefont {Medvidovi{\'c}}}\ and\ \bibinfo {author} {\bibfnamefont {G.}~\bibnamefont {Carleo}},\ }\href {https://doi.org/10.1038/s41534-021-00440-z} {\bibfield  {journal} {\bibinfo  {journal} {npj Quantum Information}\ }\textbf {\bibinfo {volume} {7}},\ \bibinfo {pages} {1} (\bibinfo {year} {2021})}\BibitemShut {NoStop}%
\bibitem [{\citenamefont {Donatella}\ \emph {et~al.}(2023)\citenamefont {Donatella}, \citenamefont {Denis}, \citenamefont {Le~Boit\'e},\ and\ \citenamefont {Ciuti}}]{Donatella2023Infidelity}%
  \BibitemOpen
  \bibfield  {author} {\bibinfo {author} {\bibfnamefont {K.}~\bibnamefont {Donatella}}, \bibinfo {author} {\bibfnamefont {Z.}~\bibnamefont {Denis}}, \bibinfo {author} {\bibfnamefont {A.}~\bibnamefont {Le~Boit\'e}},\ and\ \bibinfo {author} {\bibfnamefont {C.}~\bibnamefont {Ciuti}},\ }\href {https://doi.org/10.1103/PhysRevA.108.022210} {\bibfield  {journal} {\bibinfo  {journal} {Phys. Rev. A}\ }\textbf {\bibinfo {volume} {108}},\ \bibinfo {pages} {022210} (\bibinfo {year} {2023})}\BibitemShut {NoStop}%
\bibitem [{\citenamefont {Gravina}\ \emph {et~al.}(2024)\citenamefont {Gravina}, \citenamefont {Savona},\ and\ \citenamefont {Vicentini}}]{Gravina2024ptVMC}%
  \BibitemOpen
  \bibfield  {author} {\bibinfo {author} {\bibfnamefont {L.}~\bibnamefont {Gravina}}, \bibinfo {author} {\bibfnamefont {V.}~\bibnamefont {Savona}},\ and\ \bibinfo {author} {\bibfnamefont {F.}~\bibnamefont {Vicentini}},\ }\href@noop {} {\bibfield  {journal} {\bibinfo  {journal} {arXiv preprint}\ } (\bibinfo {year} {2024})},\ \Eprint {https://arxiv.org/abs/arXiv:2410.10720} {arXiv:2410.10720} \BibitemShut {NoStop}%
\bibitem [{\citenamefont {Carleo}\ \emph {et~al.}(2019)\citenamefont {Carleo} \emph {et~al.}}]{netket2}%
  \BibitemOpen
  \bibfield  {author} {\bibinfo {author} {\bibfnamefont {G.}~\bibnamefont {Carleo}} \emph {et~al.},\ }\href {https://doi.org/10.1016/j.softx.2019.100311} {\bibfield  {journal} {\bibinfo  {journal} {SoftwareX}\ ,\ \bibinfo {pages} {100311}} (\bibinfo {year} {2019})}\BibitemShut {NoStop}%
\bibitem [{\citenamefont {Vicentini}\ \emph {et~al.}(2022)\citenamefont {Vicentini} \emph {et~al.}}]{vicentini2022netket}%
  \BibitemOpen
  \bibfield  {author} {\bibinfo {author} {\bibfnamefont {F.}~\bibnamefont {Vicentini}} \emph {et~al.},\ }\href {https://doi.org/10.21468/SciPostPhysCodeb.7} {\bibfield  {journal} {\bibinfo  {journal} {SciPost Phys. Codebases}\ ,\ \bibinfo {pages} {7}} (\bibinfo {year} {2022})}\BibitemShut {NoStop}%
\bibitem [{\citenamefont {Westerhout}(2020)}]{spined}%
  \BibitemOpen
  \bibfield  {author} {\bibinfo {author} {\bibfnamefont {T.}~\bibnamefont {Westerhout}},\ }\href {https://github.com/twesterhout/spin-ed} {\emph {\bibinfo {title} {SpinED}}} (\bibinfo {year} {2020})\BibitemShut {NoStop}%
\bibitem [{\citenamefont {Ma}\ \emph {et~al.}(2025)\citenamefont {Ma}, \citenamefont {Liu}, \citenamefont {Li}, \citenamefont {Zhang}, \citenamefont {Duan}, \citenamefont {Wu},\ and\ \citenamefont {Deng}}]{ma2025solving}%
  \BibitemOpen
  \bibfield  {author} {\bibinfo {author} {\bibfnamefont {Y.}~\bibnamefont {Ma}}, \bibinfo {author} {\bibfnamefont {C.}~\bibnamefont {Liu}}, \bibinfo {author} {\bibfnamefont {W.}~\bibnamefont {Li}}, \bibinfo {author} {\bibfnamefont {S.-Y.}\ \bibnamefont {Zhang}}, \bibinfo {author} {\bibfnamefont {L.-M.}\ \bibnamefont {Duan}}, \bibinfo {author} {\bibfnamefont {Y.}~\bibnamefont {Wu}},\ and\ \bibinfo {author} {\bibfnamefont {D.-L.}\ \bibnamefont {Deng}},\ }\href@noop {} {\bibfield  {journal} {\bibinfo  {journal} {arXiv preprint}\ } (\bibinfo {year} {2025})},\ \Eprint {https://arxiv.org/abs/arXiv:2506.08594} {arXiv:2506.08594} \BibitemShut {NoStop}%
\end{thebibliography}%

\onecolumngrid

\appendix

\section*{Appendix}

\subsection{Matrix Minors}
Here we report key linear algebra identities that are used in the next sections of the Appendix for deriving the determinant sampling averages.
These identities describe how to compute matrix minors, namely determinants of square submatrices, with their complementary minors of the inverse matrix. 
For an $N \times N$ matrix $W$ and a collection rows $I=(i_{1},\ldots, i_{M} )$  and columns  $J=(j_{1},\ldots, j_{M} )$, let $W_{I \, J}$ ( $W_{ \overline{I} \, \overline{J} }$  ) denote the $M \times M$ ($(N-M) \times (N-M)$) submatrices given by selecting (removing) rows $I$ and columns $J$. Additionally, $W_{I \, J}$ is given so that the selected rows and columns are permuted to match orderings as given by $I$ and $J$ respectively.
The minors of $W$ for removing rows $I$ and columns $J$ can be expressed in terms of the minors of $W$ for including rows $J$ and columns $I$ and the determinant of $W$ as
\begin{equation}
        \det [ W_{\overline{I} \,\overline{J} } ] = \pm 
        \det [  (W^{-1})_{J \, I} ]  \, \det [ W ] .
        \label{eq:matrix minors general}
\end{equation}

When rows and columns are normally ordered as $i_1 < i_2 < \ldots $ and $j_1 < j_2 < \ldots $, the $\pm$ sign is given by $(-1)^{i_1 + j_1 + \ldots + i_M + j_M} $.
For general orderings of $I$ and $J$, as permutations of the normal orderings,  the $\pm$ sign is given by the signs of these two permutations.
In the following sections of the Appendix, the identity will only be applied for the lowest cases of $M=1$ or $M=2$.
In these scenarios, the identity can be expressed as
\begin{equation}
\begin{aligned}
        \det [ W_{\overline{i_1} \,\overline{j_1} } ] &= (-1)^{i_1 +j_1} \left( W^{-1} \right)_{j_1 \, i_1}  \,  \det [ W ] \\
        \det [ W_{\overline{(i_1 , i_2)} \,\overline{(j_1 , j_2)} } ] &= (-1)^{i_1 +j_1 +i'_2 + j'_2} \left( W^{-1} \right)_{j_1 \, i_1}  \left(  \left(W_{\overline{i_1} \, \overline{j_1} } \right)^{-1} \right)_{j'_2 \, i'_2} \, \det [ W ],
        \label{eq:matrix minors}
    \end{aligned}
\end{equation}
where the primed indices denote the new relative indices for the corresponding unprimed indices after removing other rows and columns.
Therefore, we have that $(W_{\overline{i_1}\overline{j_1}})_{i'_2 j'_2} =W_{i_2 j_2}$ which gives $i'_2 = i_2$ ($j'_2 = j_2$) if $i_2 < i_1$ ($j_2 < j_1$) and  $i'_2 = i_2 -1$ ($j'_2 = j_2 -1$) if $i_2 > i_1$ ($j_2 > j_1$).

In the derivations of the determinant sampling averages,~\cref{eq:matrix minors} is applied repeatedly to the $N \times N$ overlap matrices $\Phi(S) = \lbb S | \Phi \rbb$ between the sampled $N$-tuples $S$ and the variational basis states $\Phi$. 
In this case of overlap matrices, the sub-matrix $(\Phi(S))_{\overline{I} \, \overline{J} }$ is then equivalent to the overlap matrix $\Phi_{\overline{J}} (S_{\overline{I}} )=\lbb S_{\overline{I}} | \Phi_{\overline{J}} \rbb$, where $S_{\overline{I}}$ and $\Phi_{\overline{J}}$ denote the ($N-M$)-tuples given by removing the states $s_{i_1},\ldots,s_{i_M}$ from $S$ and the states $\phi_{j_1},\ldots,\phi_{j_M}$ from $\Phi$ respectively.
Importantly, this notation signifies that the minors $\det[(\Phi(S))_{\overline{I} \overline{J} }]=\det[\Phi_{\overline{J}} (S_{\overline{I}} )]$ are functionally independent of the $M$ states $S_{I} =(s_{i_1},\ldots,s_{i_M})$. 
The application of the identity~\cref{eq:matrix minors} will allow the summations over $S$ to be factorized separately over $S_{I}$ and $S_{\overline{I}}$, with the ladder terms given by a minor and a complex conjugate minor. From~\cref{eq:wedge_inner_det}, a series of minors and conjugate minors of $\Phi(S)$ results in a corresponding Gram matrix minor as
\begin{equation}
    \begin{aligned}
        \sum_{S \in \mathcal{S}^{N-M}} \det[\Phi_{I}(S) ]^* \det[\Phi_{J}(S) ] = (N-M)! \, \det[ \,(G(\Phi))_{\overline{I} \overline{J} } \, ].
        \label{eq:Sum of Matrix Minors}
    \end{aligned}
\end{equation}

\subsection{Determinant Sampling Averages: One Local Matrix}
\label{sec:one_local_matrix}
Let us consider a general normalized overlap of the type $\bra{\phi} \ket{a} / \bra{\phi} \ket{\phi}$. 
This includes the case of an expectation value when $\ket{a} = \hat{A} \ket{\phi}$, for a given operator $\hat{A}$. 
For single-state VMC, the derivation of the average of the local estimator is straightforward.
The $\phi(s) = \langle s | \phi \rangle $ from the probabilities cancels out the $\phi(s)$ in the denominator of the local estimator, leaving the multiplication between $\langle s | a \rangle $ and $\langle \phi | s \rangle$ in the numerator.
Then, the summation over $s$ results in the overlap $\langle \phi | a \rangle $ normalized by $\langle \phi | \phi \rangle$.
Now, for the determinant sampling averages of local matrix estimators, the $\phi(s)$ from the probabilities become $\det[\Phi(S)]$ and the $\phi(s)$ in the denominator become the inverse matrix $\Phi^{-1}(S)$. 
The local matrices are constructed as $\widetilde{A}(S) = \Phi^{-1}(S) \cdot A(S)$, where  $A(S) = \lbb S | A \rbb$.
Local operator matrices correspond to taking $|A\rbb = \hat{A} |\Phi \rbb$, while local ratio matrices correspond to taking $|A\rbb = | \Psi \rbb$. 
The average of local matrices $\Ex[A(S)] = \lbb \widetilde{\Phi}|A \rbb$ can be decomposed to give the more general averages: 
\begin{equation}
\label{eq:One Local Mat}
        \displaystyle \Ex \left[  \left(\Phi^{-1}(S) \right)_{i k} A_{k j}(S)  \right] = \frac{1 }{N } \langle \widetilde{\phi}_i | a_j \rangle.
\end{equation}

The key component of the derivation is that, instead of simply canceling, the identity~\cref{eq:matrix minors} is applied transforming matrix determinants to matrix minors.
Firstly, conjugated terms need to be inserted to match the unconjugated terms from the local matrices.
In order to also apply the identity~\cref{eq:matrix minors} to  $\det[\Phi(S)]^*$ from the probabilities, conjugated $\Phi^{-1}$ matrix elements must be added.
Then, the overlaps $A_{k j}(S) = \langle s_k | a_j \rangle $ from the local matrices must be paired with added overlaps $\Phi^*_{k \ell}(S) =\langle \phi_\ell | s_k \rangle$. Like for the single state VMC case, this will eventually allow the summations over $s_k$ to result in physical overlaps $\langle \phi_\ell | a_j \rangle $.
Both of these needed additional terms are given by inserting the identity $1 =   (\Phi(S) \cdot \Phi^{-1}(S))_{kk}^* $ into each term of the series. 
So, we obtain:
\begin{equation*}
    \begin{aligned}
        \displaystyle \Ex \left[  \left(\Phi^{-1}(S) \right)_{i k} A_{k j}(S)  \right] &= \frac{1}{N! \det [G]}  \sum_{S \in \mathcal{S}^N} |\det [\Phi(S)] |^2 \left(\Phi^{-1}(S) \right)_{i k} A_{k j}(S) \\
        &= \frac{1}{N! \det [G]}  \sum_{S \in \mathcal{S}^N} \Biggl(  `` \quad " \Biggl)  \left( \sum_{\ell = 1}^{N} \Phi_{k \ell}(S)(\Phi^{-1}(S))_{\ell k}   \right)^* \cdot \,   \\
    \end{aligned}
\end{equation*}

In the preceding and all subsequent equations, the averages are evaluated over the distribution $P_\Phi(S)$ ~\cref{eq:det_samp}.
Pulling out the inserted summation over $\ell$, the conjugated terms of the series now match the unconjugated terms.
Then, after applying the identity~\cref{eq:matrix minors} and pulling out the overall sign, both sets of terms are then given by a matrix minor of $\Phi(S)$ corresponding to removing row $k$ and an overlap with the state $s_k$.
The series over the $N$-tuples $S$ can then be factorized into sum of the overlaps of $s_k$ and sum of the matrix minors over the other $N-1$ states $S_{\overline{k}}$.
Thus, we get:

\begin{equation*}
    \begin{aligned}
        &= \frac{1}{N! \det [G]} \sum_{\ell =1}^N \sum_{S \in \mathcal{S}^N} 
        \biggl( \det [\Phi(S)] \, (\Phi^{-1}(S))_{\ell k} \, \Phi_{k \ell}(S)    \biggl)^* \,  \biggl( \det [\Phi(S)] \, (\Phi^{-1}(S))_{i k} \,  A_{k j} (S)   \biggl)\\
        &= \frac{ 1}{N! \det [G]} \sum_{\ell =1}^N \sum_{S \in \mathcal{S}^N} 
        \biggl( (-1)^{k + \ell} \det [\left( \, \Phi(S) \, \right)_{\overline{k} \overline{\ell} }  ] \, \Phi_{k+ \ell}(S)    \biggl)^* \,  \biggl((-1)^{k + i} \det [\left( \, \Phi(S) \, \right)_{\overline{k} \overline{i}}] \,   A_{k j} (S)   \biggl)\\
        &= \frac{(-1)^{i + \ell}}{N! \det [G]} \sum_{\ell =1}^N \Biggl( \sum_{S_{\overline{k}} \in \mathcal{S}^{N-1}}  \det[\Phi_{\overline{\ell}}(S_{\overline{k} }) ] ^* \, \det[\Phi_{\overline{i}}(S_{\overline{k} }) ] \Biggl)
        \Biggl( \sum_{s_k \in \mathcal{S}} \langle \phi_{\ell}|s_k \rangle \langle s_k | a_j \rangle  \Biggl).
    \end{aligned}
\end{equation*}

The sums of $s_k$ overlaps result in the overlaps $\langle \phi_\ell | a_j \rangle$ like for single state VMC averages, while the sums of matrix minors of $\Phi(S)$ over $S_{\overline{k}}$ from~\cref{eq:Sum of Matrix Minors} result in the matrix minors of the Gram matrix $G(
\Phi)$ with a factor of $(N-1)!$.
Then, the matrix minor identity~\cref{eq:matrix minors} is applied once again, but in reverse and in terms of the Gram matrix.
So, the resulting matrix minors, the determinant from the partition function, and the composite sign from the original double application of the identity are transformed into inverse matrix elements.
This leads to:

\begin{equation*}
    \begin{aligned}
        &= \frac{(-1)^{i + \ell}  }{N! \det [G]} \sum_{\ell =1}^N \biggl(  (N-1)! \, \det[G_{\overline{\ell}{i}}]
         \biggl)
        \biggl( \langle \phi_\ell | a_j \rangle  \biggl) \\
        &=  \frac{1 }{N } \sum_{\ell =1}^N \left( \frac{ (-1)^{i + \ell} \det [G_{\overline{\ell} \overline{i}} ] }{\det [G] } \right) \,   \langle \phi_{\ell} | a_j \rangle = \frac{1 }{N } \sum_{\ell =1}^N (G^{-1})_{i \ell} \langle \phi_{\ell} | a_j \rangle  \\
        &=   \frac{1 }{N } \langle \widetilde{\phi}_i | a_j \rangle.
    \end{aligned}
\end{equation*}

The inserted summation over $\ell$ from the first step transforms the inverse Gram matrix elements and the $\Phi$ basis overlaps with $a_j$ into a single dual basis overlap with $a_j$.
The normalization by a factor $1/N$ arises from the factor of $(N-1)!$ from the sums of matrix minors and the factor of $N!$ from the partition function. 

\subsection{Determinant Sampling Averages: Two Local Matrices}
\label{sec:two_local_matrices}
The averages for two local matrices can be expressed most clearly in terms of their sampling covariances.
For the most general case of two ordered sets of states $A=(a_1,\ldots,a_N)$  and $B=(b_1,\ldots,b_N)$ with local matrices $\widetilde{A}(S) = \Phi^{-1}(S)\cdot A(S)$ and $\widetilde{B}(S) = \Phi^{-1}(S)\cdot B(S)$ respectively, their covariance expressed in terms of tensor products $\Ex[\widetilde{A}^\dagger(S) \otimes \widetilde{B}(S)] - \Ex[\widetilde{A}(S)]^\dagger \otimes \Ex[\widetilde{B}(S')] $ equates to the tensor product $G^{-1} \otimes \lbb A |\hat{P}_{\mathcal{V}^\perp }| B \rbb $, but with respective indices permuted.
Therefore, the covariance of each pair of individual local matrix elements specifically equates to
\begin{equation}
    \Ex \left[ \widetilde{A}_{i_1 j_1 }^* (S) \, \widetilde{B}_{i_2 j_2 }(S)  \right] - \Ex \left[ \widetilde{A}_{i_1 j_1 }^* (S)\right] \Ex \left[  \widetilde{B}_{i_2 j_2 }(S')  \right] = (G^{-1})_{i_2 i_1 } \, \langle a_{j_1}|\hat{P}_{\mathcal{V}^\perp }| b_{j_2} \rangle
    \label{eq:Two Local Mat Cov}.
\end{equation}

As for operator expectation matrices, the sampling expression for operator covariance matrices~\cref{eq:cov_mat_samp} among operators $\hat{A}$ and $\hat{B}$ can be obtained for the case where $|A\rbb = \hat{A}|\Phi \rbb$ and similarly $|B\rbb = \hat{B}| \Phi \rbb$.
Contracting over the indices $i_1=j_1$ in the above equation then yields the matrix element $\widetilde{\Sigma}^{(AB)}_{i_2 j_2}$. 

Even more general sampling averages can be obtained for each of the individual terms after decomposing the local matrices as $\widetilde{A}_{i_1 j_1}(S) = \sum_{k_1} (\Phi^{-1}(S))_{i_1 k_1} A_{k_1 j_1}(S)$ and  $\widetilde{B}_{i_2 j_2}(S) = \sum_{k_2} (\Phi^{-1}(S))_{i_2 k_2} A_{k_2 j_2}(S)$. The value of the averages for each combination of indices $k_1$ and $k_2$, corresponding to the overlaps with the individual states $s_{k_1}$ and $s_{k_2}$ from the sampled $N$-tuples $S$ respectively, is dependent on whether $k_1 = k_2$ or $k_1 \neq k_2$.
Therefore, we obtain:
\begin{equation*}
    \begin{aligned}
        &\Ex \left[ \left( (\Phi^{-1}(S))_{i_1 k_1} A_{k_1 j_1}(S) \right)^* \, \left( (\Phi^{-1}(S))_{i_2 k_2} B_{k_2 j_2}(S) \right) \right] =\\
        & \frac{1}{N} (G^{-1})_{i_2 i_1} \langle a_{j_1}| b_{j_2} \rangle &\text{ for } k_1 = k_2 \\
        &\frac{1}{N(N-1)} \left( \langle a_{j_1}| \widetilde{\phi}_{i_1} \rangle \langle \widetilde{\phi}_{i_2} | b_{j_2} \rangle - (G^{-1})_{i_2 i_1} \langle a_{j_1}|\hat{P}_{\mathcal{V}}| b_{j_2} \rangle \right) &\text{ for } k_1 \neq k_2
    \end{aligned}
\end{equation*}

The terms for the average of the $N$ $k_1 = k_2$ combinations and the $N(N-1)$ $k_1 \neq k_2$ combinations are scaled by corresponding factors $1/N$ and $1/N(N-1)$ respectively.
So, the average for the local matrix elements is the sum of the two terms for each case. The local matrix covariance~\cref{eq:Two Local Mat Cov} is thus formed combining the $k_1 = k_2$ term, given by the overlap without the complementary projector $\hat{P}_{\mathcal{V}^\perp} = 1 -\hat{P}_{\mathcal{V}}$ and the $k_1 \neq k_2$ and $k_1 \neq k_2$ term, given by the negative of the overlap with the projector $\hat{P}_{\mathcal{V}}$ plus the subtracted averages $\Ex[\widetilde{A}_{i_1 j_1}^*(S)] = \langle a_{j_1}| \widetilde{\phi}_{i_1} \rangle$ times 
$\Ex[\widetilde{B}_{i_2 j_2}(S)] =\langle \widetilde{\phi}_{i_2} | b_{j_2} \rangle$.

The derivation of the $k_1 = k_2 = k$ case follows essentially the same steps as the derivation for the single local matrix averages~\cref{eq:One Local Mat} in terms for $B$ and indices $i_2 $ and $ j_2$.
The only difference is that instead of inserting the identity $1 =(\Phi^{-1}(S) \Phi(S))_{kk}^* $, those series of terms are replaced by the single term $(\Phi^{-1}(S))_{i_1 k} A_{k j_1}(S) )^*$.
This leads to:
\begin{equation*}
    \begin{aligned}
        &\Ex \left[ \left( (\Phi^{-1}(S))_{i_1 k} A_{k j_1}(S) \right)^* \, \left( (\Phi^{-1}(S))_{i_2 k} B_{k j_2}(S) \right) \right] \\
        &= \frac{1}{N! \det[G]} \sum_{S \in \mathcal{S}^N} \left( \det[\Phi(S)] (\Phi^{-1}(S))_{i_1 k} A_{k j_1}(S) \right)^* \, \left( \det[\Phi(S)](\Phi^{-1}(S))_{i_2 k} B_{k j_2}(S) \right) \\
        &= \frac{(-1)^{i_1 +i_2}}{N! \det[G]} \left( \sum_{S_{\overline{k}} \in \mathcal{S}^{N-1}} 
        \det[\Phi_{\overline{i_1}}(S_{\overline{k}})]^* \, \det[ \Phi_{\overline{i_2}}(S_{\overline{k}})] \right) \, \left( \sum_{s_k \in \mathcal{S} } \langle a_{j_1}|s_k \rangle \langle s_k | b_{j_2} \rangle \right) \\
        &= \frac{(N-1)!}{N!} \left( \frac{ (-1)^{i_1 +i_2} \det[G_{\overline{i_1} \overline{i_2} } ]}{ \det[G]} \right) \langle a_{j_1}|b_{j_2} \rangle = \frac{1}{N} (G^{-1})_{i_2 i_1} \langle a_{j_1}| b_{j_2} \rangle.
    \end{aligned}
\end{equation*}

For the $k_1 \neq k_2 $ case, the derivation is much more involved than before, as the summations over $S$ now have to be factorized separately over $s_{k_1}$,$s_{k_2}$, and the other $N-2$ states $S_{\overline{(k_1,k_2)} }$.
The matrix minor identity~\cref{eq:matrix minors} then needs to be applied for the complicated case of removing two rows ($k_1$ and $k_2$) and two columns.
Also, two overlaps $\langle s_{k_1} | \phi_{\ell_2} \rangle$ and $\langle \phi_{\ell_1} | s_{k_2}\rangle$ need to be added to match respectively the overlaps $A_{k_1 j_1}^*(S) = \langle a_{j_1} | s_{k_1} \rangle$ and $B_{k_2 j_2}(S) = \langle s_{k_2}| b_{j_2} \rangle$ from two local matrices.
Together, this is achieved by doubly inserting the identities $1=( (\Phi(S))_{\overline{k_1} \overline{i_1} }\cdot ((\Phi(S))_{\overline{k_1} \overline{i_1} })^{-1}   )_{k'_{2} k'_{2} }^* $ and $1=( (\Phi(S))_{\overline{k_2} \overline{i_2} }\cdot ((\Phi(S))_{\overline{k_2} \overline{i_2} })^{-1}   )_{k'_{1} k'_{1} } $ into each term of the series.
Again, the primed indices denote what the unprimed indices become after the removal of another row/column index, so that $k'_1 = k_1$ and $k'_2 =k_2 -1$ if $k_1 < k_2$ and $k'_1 = k_1-1$ and $k'_2 =k_2$ if $k_1 > k_2$.
Thus, we can write:

\begin{equation*}
    \begin{aligned}
        &\Ex \left[ \left( (\Phi^{-1}(S))_{i_1 k_1} A_{k_1 j_1}(S) \right)^* \, \left( (\Phi^{-1}(S))_{i_2 k_2} B_{k_2 j_2}(S) \right) \right] \\
        &= \frac{1}{N! \det[G]} \sum_{S \in \mathcal{S}^N} \biggl( \det[\Phi(S)] (\Phi^{-1}(S))_{i_1 k_1} A_{k_1 j_1}(S) \biggl)^* \, \biggl( \det[\Phi(S)](\Phi^{-1}(S))_{i_2 k_2} B_{k_2 j_2}(S) \biggl) \\
        &= \frac{1}{N! \det[G]} \sum_{S \in \mathcal{S}^N} \biggl( `` \quad " \biggl)^*\biggl( `` \quad " \biggl) \biggl( \sum_{\ell_1 \neq i_1} \Phi_{k_2 \ell_1}(S) \Bigl( ((\Phi(S))_{\overline{k_1 i_1} } )^{-1} \Bigl)_{\ell'_1 k'_2} \biggl)^*
        \biggl( \sum_{\ell_2 \neq i_2} \Phi_{k_1 \ell_2}(S) \Bigl( ((\Phi(S))_{\overline{k_2 i_2} } )^{-1} \Bigl)_{\ell'_2 k'_1} \biggl)\\
        &= \frac{ \mp 1 }{N! \det[G]} \sum_{\substack{\ell_1 \neq i_1 \\ \ell_2 \neq i_2}} \sum_{S \in \mathcal{S}^N} \biggl(  \det[ (\Phi(S))_{\overline{(k_1,k_2)} \overline{(i_1,\ell_1)} } ] A_{k_1 j_1 }(S) \Phi_{k_2 \ell_1}(S)\biggl)^*
        \biggl(  \det[ (\Phi(S))_{\overline{(k_2,k_1)} \overline{(i_2,\ell_2)} } ] B_{k_2 j_2 }(S) \Phi_{k_1 \ell_2}(S)\biggl).
    \end{aligned}
\end{equation*}

The inserted summations over $\ell_1$ and $\ell_2$ are pulled out, and the indices $\ell'_1$ and $\ell'_2$ similarly denote the sub matrix indices for $\ell_1$ and $\ell_2$  after removing columns $i_1$ and $i_2$ respectively.
As before, the conjugated and unconjugated terms are grouped and the identity~\cref{eq:matrix minors} is applied to both.
Each is then given individually by an overlap with $s_{k_1}$ an overlap with $s_{k_2}$ and a matrix minor of $\Phi(S)$ independent of both $s_{k_1}$ and $s_{k_2}$.
Then, the sums over the two overlaps with $s_{k_1}$ and the two overlaps with $s_{k_2}$ can both be factored out separately, with each resulting in a physical overlap.
As before, the sum of the matrix minors over the remaining $N-2$ states results in Gram matrix minors from~\cref{eq:Sum of Matrix Minors} but now for $(N-2)\times (N-2)$ submatrices. The overall sign in the last line, given by the product of two signs $(-1)^{i_1+k_1 + \ell'_1 +k'_2}$ and $(-1)^{ i_2 +k_2 +\ell'_2 +k'_1}$ from the respective applications of the matrix minor identity, is denoted as $\mp1$ as equals the negative of the $\pm1 =(-1)^{ i_1 +i_2 +\ell'_1 +\ell'_2}$ sign that will be eventually absorbed when the matrix minor identity is applied again on the Gram matrix minors, leaving an overall negative in the final result.
Hence, it follows that:

\begin{equation*}
    \begin{aligned}
        &= \frac{\mp1}{N! \det[G]} \sum_{\ell_1 \neq i_1} \sum_{\ell_2 \neq i_2} \left( 
        \sum_{S_{\overline{(k_1,k_2)}} } \det[\Phi_{\overline{(i_1,\ell_1)} }\left(S_{\overline{(k_1,k_2)}}\right)]^*\det[\Phi_{\overline{(i_2,\ell_2)} }\left(S_{\overline{(k_1,k_2)}}\right)]\right) \cdot \\
        &\cdot \Biggl( \sum_{s_{k_1}} \langle a_{j_1} |s_{k_1} \rangle \langle s_{k_1}| \phi_{\ell_{2}} \rangle \Biggl) \Biggl( \sum_{s_{k_2}} \langle \phi_{\ell_{1}}|s_{k_2} \rangle \langle s_{k_2}| b_{j_2} \rangle
        \Biggl) \\
        &= \frac{\mp1}{N! \det[G]} \sum_{\ell_1 \neq i_1} \sum_{\ell_2 \neq i_2} \biggl((N-2)! \, \det[G_{\overline{(i_1,\ell_1)} \overline{(i_2,\ell_2)} }] \biggl) \langle a_{j_1}|\phi_{\ell_2} \rangle \langle \phi_{\ell_1} | b_{j_2} \rangle \\
        &= \frac{-1}{N(N -1)} \sum_{\ell_1 \neq i_1} \sum_{\ell_2 \neq i_2} \left(\frac{\pm \det[G_{\overline{(i_1,\ell_1)} \overline{(i_2,\ell_2)} }]}{\det[G] } \right) \langle a_{j_1}|\phi_{\ell_2} \rangle \langle \phi_{\ell_1} | b_{j_2} \rangle \\
        &= \frac{-1}{N(N -1)} \sum_{\ell_1 =1}^N \sum_{\ell_2 =1}^N \left( \det[(G^{-1})_{(i_2,\ell_2)(i_1,\ell_1)}]\right) \langle a_{j_1}|\phi_{\ell_2} \rangle \langle \phi_{\ell_1} | b_{j_2} \rangle.
    \end{aligned}
\end{equation*}

The terms of the partition function are then incorporated. Firstly, the $N!$ with the factor of  $(N-2)!$ from the redundant orderings in the sums of matrix minors gives the overall factor $1/N(N-1)$.
The Gram matrix determinant with the resulting Gram minors and $\pm 1$ sign are transformed using the matrix minor identity into the inverse Gram matrix minors, now for  $2\times 2$ sub-matrices. In the last line, the summations over the inserted $\ell_1$ and $\ell_2$ indices are trivially extended to include $\ell_1=i_1$ and $\ell_2=i_2$ as all the added terms have matrix minors for selecting duplicate rows or columns, which are zero.
This leads to:
\begin{equation*}
    \begin{aligned}
        &=\frac{-1}{N(N -1)} \sum_{\ell_1 =1}^N \sum_{\ell_2 =1}^N \left((G^{-1})_{i_2 i_1} (G^{-1})_{\ell_2 \ell_1} - (G^{-1})_{i_2 \ell_1} (G^{-1})_{\ell_2 i_1} \right) \langle a_{j_1}|\phi_{\ell_2} \rangle \langle \phi_{\ell_1} | b_{j_2} \rangle \\
        &= \frac{1}{N(N -1)} \left(\langle a_{j_1}|\widetilde{\phi}_{i_1} \rangle \langle \widetilde{\phi}_{i_2} | b_{j_2} \rangle - (G^{-1})_{i_2 i_1}\, \langle a_{j_1}| \hat{P}_{\mathcal{V}}|b_{j_2} \rangle \right).
    \end{aligned}
\end{equation*}

Finally, the inverse Gram minors are expressed in terms of the individual inverse matrix elements. Contracting over all the $\ell_1$ and $\ell_2$ indices with the corresponding indices for $\Phi$ basis states in the overlaps again transforms them into $\Phi$ dual basis states.
One of the terms can then be expressed in terms of the orthogonal projector $\hat{P}_{\mathcal{V}} = \sum_{\ell_1} |\widetilde{\phi}_{\ell_1} \rangle \langle \phi_{\ell_1}| = \sum_{\ell_2} |\phi_{\ell_2} \rangle \langle \widetilde{\phi}_{\ell_2}|$, eventually yielding~\cref{eq:Two Local Mat Cov}.

\newpage
\subsection{Numerical values}
In the tables below we show the numerical values for the variational energies obtained in this work.

\begin{table}[htbp]
\footnotesize
\centering
\begin{tabular}{|c|c|c|c|c|c|}
\hline
$(q_x, q_y, q_{\text{sf}})$ & Data & $\ket{E_0}$ & $\ket{E_1}$ & $\ket{E_2}$ & $\ket{E_3}$ \\
\hline
$(0, 0, 0)$ & \makecell{$E/N_s$\\$E_{\text{ED}}/N_s$} & \makecell{$-0.678871(3)$\\$-0.67887215$} & \makecell{$-0.655053(4)$\\$-0.65506482$} & \makecell{$-0.603897(6)$\\$-0.60391248$} & \makecell{$-0.602881(7)$\\$-0.60291171$} \\ \hline
$(0, 0, 1)$ & \makecell{$E/N_s$\\$E_{\text{ED}}/N_s$} & \makecell{$-0.60019(2)$\\$-0.60032623$} & \makecell{$-0.60019(2)$\\$-0.60032623$} & \makecell{$-0.56901(2)$\\$-0.56928444$} & \makecell{$-0.56893(2)$\\$-0.56928444$} \\ \hline
$(1, 0, 0)$ & \makecell{$E/N_s$\\$E_{\text{ED}}/N_s$} & \makecell{$-0.616078(8)$\\$-0.61610948$} & \makecell{$-0.593592(9)$\\$-0.59366660$} & \makecell{$-0.58389(1)$\\$-0.58398204$} & \makecell{$-0.57658(1)$\\$-0.57668706$} \\ \hline
$(1, 0, 1)$ & \makecell{$E/N_s$\\$E_{\text{ED}}/N_s$} & \makecell{$-0.630660(4)$\\$-0.63066900$} & \makecell{$-0.59368(1)$\\$-0.59377745$} & \makecell{$-0.58899(1)$\\$-0.58903232$} & \makecell{$-0.58024(1)$\\$-0.58034763$} \\ \hline
$(2, 0, 0)$ & \makecell{$E/N_s$\\$E_{\text{ED}}/N_s$} & \makecell{$-0.604259(7)$\\$-0.60429704$} & \makecell{$-0.597979(7)$\\$-0.59801604$} & \makecell{$-0.587738(9)$\\$-0.58779449$} & \makecell{$-0.585182(8)$\\$-0.58523980$} \\ \hline
$(2, 0, 1)$ & \makecell{$E/N_s$\\$E_{\text{ED}}/N_s$} & \makecell{$-0.611715(6)$\\$-0.61174367$} & \makecell{$-0.59452(1)$\\$-0.59462239$} & \makecell{$-0.57791(1)$\\$-0.57803998$} & \makecell{$-0.57381(1)$\\$-0.57390355$} \\ \hline
$(3, 0, 0)$ & \makecell{$E/N_s$\\$E_{\text{ED}}/N_s$} & \makecell{$-0.608316(8)$\\$-0.60838086$} & \makecell{$-0.595396(9)$\\$-0.59545341$} & \makecell{$-0.57700(1)$\\$-0.57712574$} & \makecell{$-0.57653(1)$\\$-0.57660315$} \\ \hline
$(3, 0, 1)$ & \makecell{$E/N_s$\\$E_{\text{ED}}/N_s$} & \makecell{$-0.611494(5)$\\$-0.61151126$} & \makecell{$-0.59700(1)$\\$-0.59709159$} & \makecell{$-0.58446(1)$\\$-0.58455869$} & \makecell{$-0.57271(1)$\\$-0.57317401$} \\ \hline
$(3, 1, 0)$ & \makecell{$E/N_s$\\$E_{\text{ED}}/N_s$} & \makecell{$-0.602065(8)$\\$-0.60212294$} & \makecell{$-0.596446(7)$\\$-0.59649054$} & \makecell{$-0.583627(8)$\\$-0.58367303$} & \makecell{$-0.57770(1)$\\$-0.57803607$} \\ \hline
$(3, 1, 1)$ & \makecell{$E/N_s$\\$E_{\text{ED}}/N_s$} & \makecell{$-0.612658(8)$\\$-0.61269218$} & \makecell{$-0.59416(1)$\\$-0.59425925$} & \makecell{$-0.58031(1)$\\$-0.58047557$} & \makecell{$-0.57749(1)$\\$-0.57761003$} \\ \hline
$(3, 2, 0)$ & \makecell{$E/N_s$\\$E_{\text{ED}}/N_s$} & \makecell{$-0.613835(6)$\\$-0.61384845$} & \makecell{$-0.601308(8)$\\$-0.60134226$} & \makecell{$-0.587228(9)$\\$-0.58728711$} & \makecell{$-0.57870(1)$\\$-0.57884628$} \\ \hline
$(3, 2, 1)$ & \makecell{$E/N_s$\\$E_{\text{ED}}/N_s$} & \makecell{$-0.631525(6)$\\$-0.63155404$} & \makecell{$-0.593036(8)$\\$-0.59311947$} & \makecell{$-0.58629(1)$\\$-0.58641191$} & \makecell{$-0.57739(1)$\\$-0.57755660$} \\ \hline
$(3, 3, 0)$ & \makecell{$E/N_s$\\$E_{\text{ED}}/N_s$} & \makecell{$-0.607775(8)$\\$-0.60781960$} & \makecell{$-0.607767(8)$\\$-0.60781960$} & \makecell{$-0.58368(1)$\\$-0.58382901$} & \makecell{$-0.583287(8)$\\$-0.58382901$} \\ \hline
$(3, 3, 1)$ & \makecell{$E/N_s$\\$E_{\text{ED}}/N_s$} & \makecell{$-0.670867(4)$\\$-0.67088497$} & \makecell{$-0.631585(5)$\\$-0.63159780$} & \makecell{$-0.593471(9)$\\$-0.59352247$} & \makecell{$-0.59264(1)$\\$-0.59271219$} \\ \hline
$(2, 2, 0)$ & \makecell{$E/N_s$\\$E_{\text{ED}}/N_s$} & \makecell{$-0.60286(1)$\\$-0.60292841$} & \makecell{$-0.602159(8)$\\$-0.60220229$} & \makecell{$-0.582344(9)$\\$-0.58239949$} & \makecell{$-0.58104(1)$\\$-0.58111876$} \\ \hline
$(2, 2, 1)$ & \makecell{$E/N_s$\\$E_{\text{ED}}/N_s$} & \makecell{$-0.617353(5)$\\$-0.61737429$} & \makecell{$-0.58580(1)$\\$-0.58593399$} & \makecell{$-0.58180(1)$\\$-0.58191747$} & \makecell{$-0.58083(1)$\\$-0.58091066$} \\ \hline
$(1, 1, 0)$ & \makecell{$E/N_s$\\$E_{\text{ED}}/N_s$} & \makecell{$-0.602883(6)$\\$-0.60292387$} & \makecell{$-0.593892(8)$\\$-0.59394598$} & \makecell{$-0.587768(8)$\\$-0.58782682$} & \makecell{$-0.581072(7)$\\$-0.58111328$} \\ \hline
$(1, 1, 1)$ & \makecell{$E/N_s$\\$E_{\text{ED}}/N_s$} & \makecell{$-0.617062(5)$\\$-0.61708493$} & \makecell{$-0.59533(1)$\\$-0.59541884$} & \makecell{$-0.579754(8)$\\$-0.57982248$} & \makecell{$-0.57513(1)$\\$-0.57523824$} \\ \hline
\end{tabular}
\caption{Energy densities $E/N_s$ and corresponding exact diagonalization (ED) values for the first 4 excited states of the Heisenberg model on the $6 \times 6$ lattice for different momentum $\boldsymbol{q}=(q_x, q_y)$ and spin-flip (indexed by $q_{\text{sf}}$) symmetry sectors.
The values of $q_x$ and $q_y$ are in units of $2 \pi / L$.
}
\label{tab:energies_6x6}
\end{table}

\begin{table}[htbp]
\centering
\footnotesize
\begin{tabular}{|c|c|c|c|c|c|c|c|}
\hline
$(q_x, q_y, q_{\text{sf}})$ & Data & $\ket{E_0}$ & $\ket{E_1}$ & $\ket{E_2}$ & $\ket{E_3}$ & $\ket{E_4}$ & $\ket{E_5}$ \\
\hline
$(0, 0, 0)$ & \makecell{$E/N_s$\\$\text{V-score}$} & \makecell{$-0.671544(4)$\\$1\cdot 10^{-4}$} & \makecell{$-0.668007(8)$\\$2\cdot 10^{-4}$} & \makecell{$-0.65987(1)$\\$4\cdot 10^{-4}$} & \makecell{$-0.65345(2)$\\$8\cdot 10^{-4}$} & \makecell{$-0.65201(1)$\\$7\cdot 10^{-4}$} & \makecell{$-0.65013(1)$\\$4\cdot 10^{-4}$} \\ \hline
$(0, 0, 1)$ & \makecell{$E/N_s$\\$\text{V-score}$} & \makecell{$-0.65200(3)$\\$3\cdot 10^{-3}$} & \makecell{$-0.65195(2)$\\$3\cdot 10^{-3}$} & \makecell{$-0.64604(2)$\\$2\cdot 10^{-3}$} & \makecell{$-0.64441(2)$\\$3\cdot 10^{-3}$} & \makecell{$-0.64366(4)$\\$4\cdot 10^{-3}$} & \makecell{$-0.64198(3)$\\$4\cdot 10^{-3}$} \\ \hline
$(1, 0, 0)$ & \makecell{$E/N_s$\\$\text{V-score}$} & \makecell{$-0.65852(1)$\\$7\cdot 10^{-4}$} & \makecell{$-0.64926(1)$\\$1\cdot 10^{-3}$} & \makecell{$-0.64703(1)$\\$1\cdot 10^{-3}$} & \makecell{$-0.64626(1)$\\$7\cdot 10^{-4}$} & \makecell{$-0.64508(1)$\\$8\cdot 10^{-4}$} & \makecell{$-0.64481(2)$\\$1\cdot 10^{-3}$} \\ \hline
$(1, 0, 1)$ & \makecell{$E/N_s$\\$\text{V-score}$} & \makecell{$-0.66079(1)$\\$5\cdot 10^{-4}$} & \makecell{$-0.65445(2)$\\$1\cdot 10^{-3}$} & \makecell{$-0.64930(3)$\\$3\cdot 10^{-3}$} & \makecell{$-0.64704(2)$\\$3\cdot 10^{-3}$} & \makecell{$-0.64359(3)$\\$3\cdot 10^{-3}$} & \makecell{$-0.64266(2)$\\$2\cdot 10^{-3}$} \\ \hline
$(2, 0, 0)$ & \makecell{$E/N_s$\\$\text{V-score}$} & \makecell{$-0.65141(1)$\\$9\cdot 10^{-4}$} & \makecell{$-0.65098(1)$\\$9\cdot 10^{-4}$} & \makecell{$-0.65009(1)$\\$8\cdot 10^{-4}$} & \makecell{$-0.64488(1)$\\$2\cdot 10^{-3}$} & \makecell{$-0.64212(2)$\\$1\cdot 10^{-3}$} & \makecell{$-0.64133(2)$\\$1\cdot 10^{-3}$} \\ \hline
$(2, 0, 1)$ & \makecell{$E/N_s$\\$\text{V-score}$} & \makecell{$-0.65308(1)$\\$7\cdot 10^{-4}$} & \makecell{$-0.64749(2)$\\$2\cdot 10^{-3}$} & \makecell{$-0.64601(2)$\\$2\cdot 10^{-3}$} & \makecell{$-0.64566(2)$\\$3\cdot 10^{-3}$} & \makecell{$-0.64151(3)$\\$4\cdot 10^{-3}$} & \makecell{$-0.63955(3)$\\$4\cdot 10^{-3}$} \\ \hline
$(3, 0, 0)$ & \makecell{$E/N_s$\\$\text{V-score}$} & \makecell{$-0.64696(2)$\\$1\cdot 10^{-3}$} & \makecell{$-0.64630(1)$\\$1\cdot 10^{-3}$} & \makecell{$-0.64242(1)$\\$1\cdot 10^{-3}$} & \makecell{$-0.64149(2)$\\$2\cdot 10^{-3}$} & \makecell{$-0.63946(2)$\\$2\cdot 10^{-3}$} & \makecell{$-0.63766(1)$\\$2\cdot 10^{-3}$} \\ \hline
$(3, 0, 1)$ & \makecell{$E/N_s$\\$\text{V-score}$} & \makecell{$-0.64914(2)$\\$1\cdot 10^{-3}$} & \makecell{$-0.64451(2)$\\$4\cdot 10^{-3}$} & \makecell{$-0.64330(3)$\\$2\cdot 10^{-3}$} & \makecell{$-0.64190(2)$\\$3\cdot 10^{-3}$} & \makecell{$-0.64036(3)$\\$3\cdot 10^{-3}$} & \makecell{$-0.63950(3)$\\$4\cdot 10^{-3}$} \\ \hline
$(4, 0, 0)$ & \makecell{$E/N_s$\\$\text{V-score}$} & \makecell{$-0.64741(2)$\\$2\cdot 10^{-3}$} & \makecell{$-0.64622(2)$\\$2\cdot 10^{-3}$} & \makecell{$-0.64001(2)$\\$2\cdot 10^{-3}$} & \makecell{$-0.63931(2)$\\$2\cdot 10^{-3}$} & \makecell{$-0.63836(2)$\\$2\cdot 10^{-3}$} & \makecell{$-0.63663(2)$\\$2\cdot 10^{-3}$} \\ \hline
$(4, 0, 1)$ & \makecell{$E/N_s$\\$\text{V-score}$} & \makecell{$-0.64860(1)$\\$1\cdot 10^{-3}$} & \makecell{$-0.64609(3)$\\$3\cdot 10^{-3}$} & \makecell{$-0.64264(2)$\\$3\cdot 10^{-3}$} & \makecell{$-0.64241(2)$\\$3\cdot 10^{-3}$} & \makecell{$-0.63894(3)$\\$5\cdot 10^{-3}$} & \makecell{$-0.63610(3)$\\$5\cdot 10^{-3}$} \\ \hline
$(5, 0, 0)$ & \makecell{$E/N_s$\\$\text{V-score}$} & \makecell{$-0.64853(2)$\\$2\cdot 10^{-3}$} & \makecell{$-0.64639(2)$\\$2\cdot 10^{-3}$} & \makecell{$-0.63953(2)$\\$2\cdot 10^{-3}$} & \makecell{$-0.63904(2)$\\$3\cdot 10^{-3}$} & \makecell{$-0.63874(2)$\\$2\cdot 10^{-3}$} & \makecell{$-0.63834(2)$\\$2\cdot 10^{-3}$} \\ \hline
$(5, 0, 1)$ & \makecell{$E/N_s$\\$\text{V-score}$} & \makecell{$-0.64902(2)$\\$1\cdot 10^{-3}$} & \makecell{$-0.64698(2)$\\$4\cdot 10^{-3}$} & \makecell{$-0.64407(2)$\\$3\cdot 10^{-3}$} & \makecell{$-0.64279(3)$\\$3\cdot 10^{-3}$} & \makecell{$-0.64089(4)$\\$5\cdot 10^{-3}$} & \makecell{$-0.63643(3)$\\$4\cdot 10^{-3}$} \\ \hline
$(5, 1, 0)$ & \makecell{$E/N_s$\\$\text{V-score}$} & \makecell{$-0.64743(2)$\\$2\cdot 10^{-3}$} & \makecell{$-0.64613(2)$\\$1\cdot 10^{-3}$} & \makecell{$-0.63980(2)$\\$2\cdot 10^{-3}$} & \makecell{$-0.63909(2)$\\$2\cdot 10^{-3}$} & \makecell{$-0.63849(2)$\\$2\cdot 10^{-3}$} & \makecell{$-0.63797(3)$\\$3\cdot 10^{-3}$} \\ \hline
$(5, 1, 1)$ & \makecell{$E/N_s$\\$\text{V-score}$} & \makecell{$-0.64872(2)$\\$1\cdot 10^{-3}$} & \makecell{$-0.64563(3)$\\$4\cdot 10^{-3}$} & \makecell{$-0.64273(2)$\\$3\cdot 10^{-3}$} & \makecell{$-0.64242(3)$\\$4\cdot 10^{-3}$} & \makecell{$-0.63922(3)$\\$6\cdot 10^{-3}$} & \makecell{$-0.63751(3)$\\$6\cdot 10^{-3}$} \\ \hline
$(5, 2, 0)$ & \makecell{$E/N_s$\\$\text{V-score}$} & \makecell{$-0.64674(2)$\\$1\cdot 10^{-3}$} & \makecell{$-0.64573(2)$\\$1\cdot 10^{-3}$} & \makecell{$-0.64272(2)$\\$2\cdot 10^{-3}$} & \makecell{$-0.64218(2)$\\$2\cdot 10^{-3}$} & \makecell{$-0.64005(1)$\\$2\cdot 10^{-3}$} & \makecell{$-0.63797(2)$\\$2\cdot 10^{-3}$} \\ \hline
$(5, 2, 1)$ & \makecell{$E/N_s$\\$\text{V-score}$} & \makecell{$-0.64922(1)$\\$1\cdot 10^{-3}$} & \makecell{$-0.64440(3)$\\$5\cdot 10^{-3}$} & \makecell{$-0.64354(2)$\\$3\cdot 10^{-3}$} & \makecell{$-0.64067(3)$\\$4\cdot 10^{-3}$} & \makecell{$-0.63993(3)$\\$4\cdot 10^{-3}$} & \makecell{$-0.63903(3)$\\$5\cdot 10^{-3}$} \\ \hline
$(5, 3, 0)$ & \makecell{$E/N_s$\\$\text{V-score}$} & \makecell{$-0.65089(1)$\\$1\cdot 10^{-3}$} & \makecell{$-0.64999(1)$\\$9\cdot 10^{-4}$} & \makecell{$-0.64702(1)$\\$1\cdot 10^{-3}$} & \makecell{$-0.64301(2)$\\$1\cdot 10^{-3}$} & \makecell{$-0.64232(1)$\\$1\cdot 10^{-3}$} & \makecell{$-0.64095(2)$\\$2\cdot 10^{-3}$} \\ \hline
$(5, 3, 1)$ & \makecell{$E/N_s$\\$\text{V-score}$} & \makecell{$-0.65315(1)$\\$9\cdot 10^{-4}$} & \makecell{$-0.64994(3)$\\$3\cdot 10^{-3}$} & \makecell{$-0.64767(3)$\\$2\cdot 10^{-3}$} & \makecell{$-0.64625(3)$\\$2\cdot 10^{-3}$} & \makecell{$-0.64320(3)$\\$4\cdot 10^{-3}$} & \makecell{$-0.64007(3)$\\$4\cdot 10^{-3}$} \\ \hline
$(5, 4, 0)$ & \makecell{$E/N_s$\\$\text{V-score}$} & \makecell{$-0.65824(1)$\\$7\cdot 10^{-4}$} & \makecell{$-0.65070(1)$\\$1\cdot 10^{-3}$} & \makecell{$-0.64832(1)$\\$1\cdot 10^{-3}$} & \makecell{$-0.64616(1)$\\$7\cdot 10^{-4}$} & \makecell{$-0.64528(1)$\\$1\cdot 10^{-3}$} & \makecell{$-0.64508(1)$\\$7\cdot 10^{-4}$} \\ \hline
$(5, 4, 1)$ & \makecell{$E/N_s$\\$\text{V-score}$} & \makecell{$-0.66089(1)$\\$7\cdot 10^{-4}$} & \makecell{$-0.65504(2)$\\$2\cdot 10^{-3}$} & \makecell{$-0.64776(3)$\\$3\cdot 10^{-3}$} & \makecell{$-0.64434(2)$\\$3\cdot 10^{-3}$} & \makecell{$-0.64401(2)$\\$4\cdot 10^{-3}$} & \makecell{$-0.64281(2)$\\$3\cdot 10^{-3}$} \\ \hline
$(5, 5, 0)$ & \makecell{$E/N_s$\\$\text{V-score}$} & \makecell{$-0.65360(1)$\\$1\cdot 10^{-3}$} & \makecell{$-0.65358(2)$\\$1\cdot 10^{-3}$} & \makecell{$-0.65022(1)$\\$6\cdot 10^{-4}$} & \makecell{$-0.64849(1)$\\$6\cdot 10^{-4}$} & \makecell{$-0.64568(1)$\\$1\cdot 10^{-3}$} & \makecell{$-0.64558(2)$\\$1\cdot 10^{-3}$} \\ \hline
$(5, 5, 1)$ & \makecell{$E/N_s$\\$\text{V-score}$} & \makecell{$-0.67032(2)$\\$7\cdot 10^{-4}$} & \makecell{$-0.66442(2)$\\$1\cdot 10^{-3}$} & \makecell{$-0.65228(4)$\\$5\cdot 10^{-3}$} & \makecell{$-0.65162(2)$\\$3\cdot 10^{-3}$} & \makecell{$-0.65000(3)$\\$3\cdot 10^{-3}$} & \makecell{$-0.64441(3)$\\$4\cdot 10^{-3}$} \\ \hline
$(4, 4, 0)$ & \makecell{$E/N_s$\\$\text{V-score}$} & \makecell{$-0.65455(1)$\\$8\cdot 10^{-4}$} & \makecell{$-0.65237(1)$\\$2\cdot 10^{-3}$} & \makecell{$-0.64959(1)$\\$7\cdot 10^{-4}$} & \makecell{$-0.64478(1)$\\$1\cdot 10^{-3}$} & \makecell{$-0.64441(1)$\\$1\cdot 10^{-3}$} & \makecell{$-0.64052(2)$\\$2\cdot 10^{-3}$} \\ \hline
$(4, 4, 1)$ & \makecell{$E/N_s$\\$\text{V-score}$} & \makecell{$-0.65701(1)$\\$8\cdot 10^{-4}$} & \makecell{$-0.65129(2)$\\$2\cdot 10^{-3}$} & \makecell{$-0.65011(2)$\\$3\cdot 10^{-3}$} & \makecell{$-0.64576(2)$\\$3\cdot 10^{-3}$} & \makecell{$-0.64170(3)$\\$3\cdot 10^{-3}$} & \makecell{$-0.64086(3)$\\$4\cdot 10^{-3}$} \\ \hline
$(3, 3, 0)$ & \makecell{$E/N_s$\\$\text{V-score}$} & \makecell{$-0.64586(2)$\\$1\cdot 10^{-3}$} & \makecell{$-0.64388(1)$\\$1\cdot 10^{-3}$} & \makecell{$-0.64243(1)$\\$9\cdot 10^{-4}$} & \makecell{$-0.64159(2)$\\$1\cdot 10^{-3}$} & \makecell{$-0.64029(2)$\\$1\cdot 10^{-3}$} & \makecell{$-0.63973(2)$\\$1\cdot 10^{-3}$} \\ \hline
$(3, 3, 1)$ & \makecell{$E/N_s$\\$\text{V-score}$} & \makecell{$-0.64811(1)$\\$1\cdot 10^{-3}$} & \makecell{$-0.64243(3)$\\$3\cdot 10^{-3}$} & \makecell{$-0.64164(3)$\\$5\cdot 10^{-3}$} & \makecell{$-0.63995(2)$\\$4\cdot 10^{-3}$} & \makecell{$-0.63873(3)$\\$4\cdot 10^{-3}$} & \makecell{$-0.63714(3)$\\$4\cdot 10^{-3}$} \\ \hline
$(2, 2, 0)$ & \makecell{$E/N_s$\\$\text{V-score}$} & \makecell{$-0.64596(1)$\\$1\cdot 10^{-3}$} & \makecell{$-0.64381(1)$\\$1\cdot 10^{-3}$} & \makecell{$-0.64224(2)$\\$1\cdot 10^{-3}$} & \makecell{$-0.64196(1)$\\$1\cdot 10^{-3}$} & \makecell{$-0.64161(1)$\\$1\cdot 10^{-3}$} & \makecell{$-0.64004(1)$\\$1\cdot 10^{-3}$} \\ \hline
$(2, 2, 1)$ & \makecell{$E/N_s$\\$\text{V-score}$} & \makecell{$-0.64810(1)$\\$1\cdot 10^{-3}$} & \makecell{$-0.64240(1)$\\$3\cdot 10^{-3}$} & \makecell{$-0.64197(3)$\\$5\cdot 10^{-3}$} & \makecell{$-0.63862(3)$\\$5\cdot 10^{-3}$} & \makecell{$-0.63751(3)$\\$4\cdot 10^{-3}$} & \makecell{$-0.63647(3)$\\$5\cdot 10^{-3}$} \\ \hline
$(1, 1, 0)$ & \makecell{$E/N_s$\\$\text{V-score}$} & \makecell{$-0.65475(1)$\\$8\cdot 10^{-4}$} & \makecell{$-0.65163(1)$\\$1\cdot 10^{-3}$} & \makecell{$-0.64964(1)$\\$7\cdot 10^{-4}$} & \makecell{$-0.64331(2)$\\$2\cdot 10^{-3}$} & \makecell{$-0.64302(2)$\\$1\cdot 10^{-3}$} & \makecell{$-0.64047(2)$\\$2\cdot 10^{-3}$} \\ \hline
$(1, 1, 1)$ & \makecell{$E/N_s$\\$\text{V-score}$} & \makecell{$-0.656952(9)$\\$7\cdot 10^{-4}$} & \makecell{$-0.65094(2)$\\$2\cdot 10^{-3}$} & \makecell{$-0.65080(3)$\\$3\cdot 10^{-3}$} & \makecell{$-0.64553(3)$\\$2\cdot 10^{-3}$} & \makecell{$-0.64258(4)$\\$4\cdot 10^{-3}$} & \makecell{$-0.64171(3)$\\$4\cdot 10^{-3}$} \\ \hline
\end{tabular}
\caption{Energy densities $E/N_s$ and V-score values for the first 6 excited states of the Heisenberg model on the $10 \times 10$ lattice for different momentum $\boldsymbol{q}=(q_x, q_y)$ and spin-flip (indexed by $q_{\text{sf}}$) symmetry sectors.
The values of $q_x$ and $q_y$ are in units of $2 \pi / L$.}
\label{tab:updated_energies}
\end{table}

\twocolumngrid

\end{document}